\documentclass[aps,  
prc,                 
showpacs,            
twocolumn,           
preprintnumbers,
nofootinbib,
floatfix]            
{revtex4-1}

\usepackage{amsmath,amsthm,amssymb} 
\usepackage{amssymb}
\usepackage{epsfig}   
\usepackage{bm}       
\usepackage{graphicx} 
\usepackage{color}
\usepackage{slashed}  
\usepackage{hyperref}
\usepackage{physics}  
\usepackage{empheq}

\usepackage[usenames,dvipsnames,svgnames,table]{xcolor}
\usepackage[utf8]{inputenc}

\hypersetup{colorlinks,linkcolor=MidnightBlue,citecolor=MidnightBlue,urlcolor=MidnightBlue}
\hypersetup{breaklinks=true}
\definecolor{babypink}{rgb}{0.96, 0.76, 0.76}
\definecolor{darkpastelgreen}{rgb}{0.01, 0.75, 0.24}

\def\Dz{D^0}
\def\Dzb{\overline{D}^0}
\def\nn{\nonumber}

\def\xp{x_\perp} 

\begin{document}

\title{Directed flow of D mesons at RHIC and LHC: non-perturbative dynamics, longitudinal bulk matter asymmetry and electromagnetic fields}

\author{Lucia Oliva$^{a}$, Salvatore Plumari$^{b,c}$ and Vincenzo Greco$^{b,c}$}

\affiliation{$^{a}$ Institut f\"{u}r Theoretische Physik, Johann Wolfgang Goethe-Universit\"{a}t, Max-von-Laue-Str. 1, 60438 Frankfurt am Main, Germany}
\affiliation{$^{b}$ Department of Physics and Astronomy, University of Catania, Via S. Sofia 64, I-95123 Catania}
\affiliation{$^{c}$ INFN-Laboratori Nazionali del Sud, Via S. Sofia 62, I-95123 Catania, Italy}


\begin{abstract}
We present a study of the directed flow $v_1$ for $D$ mesons discussing both the impact of initial vorticity and electromagnetic field.
Recent studies predicted that $v_1$ for $D$ mesons is expected to be surprisingly much larger than that of light charged hadrons;
we clarify that this is due to a different mechanism leading to the formation of a directed flow with respect to the one of the bulk matter at both relativistic and non-relativistic energies. 
We point out that the very large $v_1$ for $D$ mesons can be generated only if there is a longitudinal asymmetry between the bulk matter and the charm quarks and if the latter have a
large non-perturbative interaction in the QGP medium. A quite good agreement with the data of STAR and ALICE is obtained if the diffusion coefficient able to correctly predict the $R_{AA}(p_T)$, $v_{2}(p_T)$ and $v_{3}(p_T)$ of $D$ meson is employed.
Furthermore, the mechanism for the build-up of the $v_1(y)$ is associated to a  quite small formation time that can be expected to be more sensitive to the initial high-temperature dependence of the charm diffusion coefficient.

We discuss also the splitting of $v_1$ for $D^0$ and $\bar D^0$ due to the electromagnetic field that is again much larger than the
one observed for charged particles and in agreement with the data by STAR that have however still error bars comparable with 
the splitting itself, while at LHC standard electromagnetic profile assuming a constant conductivity is not able to account for the huge splitting observed.
\end{abstract}


\maketitle

\section{Introduction}

The theoretical and experimental studies on the formation and evolution of the Quark-Gluon Plasma (QGP) in ultra-Relativistic Heavy-Ion Collisions (uRHICs) have launched a new investigation stage, in which the main discoveries of the last decades on the hot QCD matter as a nearly inviscid fluid with strongly non-perturbative behaviour and the development of collective flows lay the foundations for dealing with new interesting features, such as the intense vorticity \cite{Becattini:2007sr, Csernai:2013bqa, Becattini:2015ska, Deng:2016gyh, Jiang:2016woz, STAR:2017ckg} and electromagnetic fields \cite{Kharzeev:2007jp, Skokov:2009qp, Voronyuk:2011jd, Tuchin:2013ie, Bzdak:2011yy, Deng:2012pc, Zhong:2014sua} generated in non-central collisions.
\\
Indeed, the fraction of the orbital angular momentum of the colliding system transferred to the hot plasma manifests itself with swirls of the order of $2-3$ fm$^{-1}$ whose observable effects could manifest in the chiral vortical effect \cite{Erdmenger:2008rm, Son:2009tf, Rogachevsky:2010ys, Kharzeev:2015znc}, the spin polarization of baryons and vector mesons \cite{Liang:2004ph, Liang:2004xn, Voloshin:2004ha, Betz:2007kg, Gao:2007bc, Sheng:2019kmk, Becattini:2020ngo} and the ``wiggle'' slope in the directed flow of hadrons \cite{Snellings:1999bt, Csernai:1999nf, Bozek:2010bi, Chatterjee:2017ahy}.
Furthermore, electromagnetic fields of the order of few to tens of $m_{\pi}^2$ are produced in the overlap area mainly due to the motion of spectator charges and can be probed by observations of the chiral magnetic effect and related quantum phenomena \cite{Kharzeev:2007jp, Fukushima:2008xe, Kharzeev:2010gd, Shi:2017cpu}, the splitting in the spin polarization \cite{Becattini:2016gvu, Han:2017hdi, Guo:2019joy} and the splitting of the directed flow  (a dipole asymmetry) \cite{Gursoy:2014aka, Voronyuk:2014rna, Toneev:2016bri, Das:2016cwd, Gursoy:2018yai, Chatterjee:2018lsx, Oliva:2019kin, Sun:2020wkg, Oliva:2020mfr, Dubla:2020bdz}; more recently also a possible impact on $v_2$, $v_3$ and the average transverse momentum $<p_T>$ has been pointed out \cite{Gursoy:2018yai}. 
In the last 15 years there has been a large effort to study the interaction of heavy quarks (HQs) with the bulk medium
\cite{Greco:2003vf,vanHees:2005wb,vanHees:2007me,He:2012df,He:2014cla,Gossiaux:2009mk,Nahrgang:2014vza,Song:2015sfa,
Song:2015ykw,Cao:2016gvr,Alberico:2011zy,Alberico:2013bza,Xu:2017obm} and more recently a first determination of its interaction in terms of a drag coefficient $\gamma(T)$ or  space-diffusion coefficient has been achieved, as expounded also in several reviews of the last two years \cite{Dong:2019unq, Dong:2019byy, Zhao:2020jqu}. There is a general consensus that the interaction is non-perturbative and around the pseudo-critical temperature $T_c$ can even get close to the value predicted in the infinite coupling limit by AdS/CFT \cite{Dong:2019unq, Xu:2017obm}.
A critical analysis of the several sources of indetermination can be found in recent joint activities comparing the features of the different approaches \cite{Rapp:2018qla, Cao:2018ews, Xu:2018gux}.
Heavy quarks and antiquarks have a very short formation time with respect to that of light quarks, therefore they are probably the earliest charged particles appearing in the uRHICs created matter.
So one may expect they keep traces of both the initial stage and the subsequent evolution into a thermalized QGP.
Furthermore, due to the large masses, their thermalization time is comparable or larger than the QGP lifetime.

A surprising novel theoretical prediction has been that the heavy quarks can manifest a directed flow $v_1=\langle p_x/p_T \rangle$ that is more than one order of magnitude larger than the one of light hadrons despite their large mass \cite{Das:2016cwd, Chatterjee:2017ahy}. 
This has been predicted for both the average charm $v_1(y)$ \cite{Chatterjee:2017ahy} and for the splitting $\Delta v_1(y)$ between particle and anti-particles ($D^0$ and $\bar D^0$) \cite{Das:2016cwd}.
In Ref.~\cite{Das:2016cwd} a Langevin approach coupled to the Maxwell equations has been used to describe the effect of the intense initial electromagnetic fields on the dynamics of heavy quarks in the QGP medium. It was predicted for the first time an electromagnetically-induced splitting in the directed flow of $\Dz$ and $\Dzb$.
\\
A similar approach has been used in Ref.~\cite{Chatterjee:2017ahy, Chatterjee:2018lsx} to study the directed flow of neutral $D$ mesons considering the tilt of the fireball in the reaction plane with respect to the beam axis. The main merit is to predict a $dv_1/dy \simeq 0.04$ which is about half the one measured by STAR \cite{Adam:2019wnk}, but a successful prediction of a directed flow of $D$ mesons can be more than one order of magnitude larger than the one of light hadrons.

Recent experimental efforts on measuring the directed flow of neutral $D$ mesons at both RHIC \cite{Adam:2019wnk} and LHC energies \cite{Acharya:2019ijj} have measured flow signals much larger than the one of light hadrons and even larger than early theoretical prediction for $v_1(y)$ or comparable with them for $\Delta v_1(y)= v_1(\Dz) - v_1(\Dzb)$ that could be associated to the presence of early electromagnetic fields.
Among the $D$ mesons, the neutral $\Dz$ and $\Dzb$ are excellent particles to study the influence of the electromagnetic fields in the QGP phase, since it is natural to expect that the influence on them of the electromagnetic fields could have origin only in the deconfined phase. 

We aim at investigating the effect of both the initial vorticity and electromagnetic fields on the directed flow of $\Dz$ and $\Dzb$ in comparison to the recent experimental measurements by STAR \cite{Adam:2019wnk} and ALICE \cite{Acharya:2019ijj} collaborations, once the longitudinal distribution of the bulk is fixed to reproduce correctly the light hadron $v_1$ and the charm-bulk interaction is the one able to describe with good accuracy the $D$ mesons observables. 
The study we present here is performed by the relativistic Boltzmann transport equation for heavy quarks developed to describe the elliptic flow $v_2$ and the nuclear modification factor $R_{AA}$ of $D$ mesons at both RHIC and LHC energies; hence it will be used as standard bulk-charm interaction the one corresponding to the determination of the non-perturbative $2\pi T D_s$ coefficient \cite{Scardina:2017ipo, Dong:2019unq}.
Moreover, we have modified our standard boost-invariant initial conditions breaking the forward-backward symmetry by means of a tilted distribution of the fireball which leads to a finite directed flow of charged particles in agreement with the experimental data.
Furthermore, with respect to the first prediction on the splitting $\Delta v_1(y)$ \cite{Das:2016cwd} at LHC energy, we include the bulk vorticity and present prediction also and mainly for RHIC energy.
We find a satisfying agreement with experimental data for both $v_1(D^0)$ and $v_1(\bar D^0)$ predicting a $dv_1/dy \simeq 0.07$ and for their difference $\Delta v_1$ finding  $d\Delta v_1/dy \simeq 0.01$  at RHIC energy.
A main part is dedicated to clarify the origin of the large $v_1$ of $D$ mesons pointing out that its origin is different from the one of the bulk QGP and that its large value manifest only due to their non-pertubative interaction with the bulk QGP matter.

The article is organized as follows.
In Sec.~\ref{transport} we introduce our transport approach, focusing on our description of the heavy quarks dynamics and the electromagnetic fields.
In Sec.~\ref{ic_vort} we explain in detail our modified initialization of the tilted fireball, while in Sec.~\ref{vortQGP} we discuss the space-time evolution of the QGP vorticity profiles.
Our results on the directed flow of neutral $D$ mesons are presented in Sec.~\ref{RHICresults} and Sec.~\ref{LHCresults} for RHIC and LHC energies respectively. Finally, we draw our conclusions.

\section{Charm quarks transport equation in the electromagnetic field}
\label{transport}

We use a relativistic transport code developed to perform studies of the dynamics of heavy-ion
collisions at both RHIC and LHC energies and different collision systems \cite{Plumari:2012ep, Ruggieri:2013bda, Scardina:2013nua, Ruggieri:2013ova, Puglisi:2014sha, Plumari:2015sia, Plumari:2015cfa, Scardina:2017ipo, Plumari:2019gwq, Sun:2019gxg}.
The dynamical evolution of gluons and light quarks as well as of charm quarks in the QGP bulk medium is described by the Relativistic Boltzmann Transport equations given by
\begin{eqnarray}
\left[ p_{\mu}\partial_{x}^{\mu} + q_j F_{\mu\nu}(x)p^{\nu}\partial_{p}^{\mu}\right]f_{j}(x,p) &=& {\cal C}[f_{j},f_{k}](x,p) \quad \label{eq:RBTeqQGP} \\
\left[ p_{\mu}\partial_{x}^{\mu} + q_Q F_{\mu\nu}(x)p^{\nu}\partial_{p}^{\mu}\right]f_{Q}(x,p) &=& {\cal C}[f_{j},f_{k},f_{Q}](x,p) \quad \label{eq:RBTeqHQ}
\end{eqnarray}
where $f_{j,k}(x,p)$ is the phase-space one-body distribution function of the parton $j,k=g,q,\overline{q}$ (gluon, quark or antiquark) and $f_Q(x,p)$ is the phase-space one-body distribution function of the heavy quark $Q=c,\overline{c}$ (charm or anticharm);
on the left-hand sides $F_{\mu\nu}$ is the electromagnetic strength tensor and on the right-hand sides ${\cal{C}}={\cal{C}}_{22}$ represents the relativistic collision integral accounting for $2\rightarrow2$ scattering processes.

For the bulk composed by quarks and gluons the evolution is described by Eqs.~\eqref{eq:RBTeqQGP} and in this paper we assume that it is independent of $f_{Q}$.
In the collision integral ${\cal C}[f_j,f_k]$ the total cross section is determined in order to keep fixed the ratio $\eta/s=1/(4\pi)$ during the evolution of the QGP,
see Refs~\cite{Plumari:2019gwq, Plumari:2015cfa, Ruggieri:2013ova, Plumari:2012xz} for more details.
However, this is equivalent to simulate the dynamical evolution of a fluid with specified $\eta/s$ by means of the Boltzmann equation.
In the present paper we have employed a bulk with massive quarks and gluons in the framework of a Quasi-Particle Model (QPM) \cite{Plumari:2011mk} where quarks and gluons are dressed with thermal masses $m_{g,q}(T)\propto g(T)\,T$ and the $T$-dependence of the coupling $g(T)$ is tuned to lattice QCD thermodynamics \cite{Borsanyi:2010cj}.
In this approach the dynamical evolution of the system has approximately the lattice QCD equation of state, with a softening of the equation of state consisting in a decreasing speed of sound approaching the cross-over region \cite{Borsanyi:2010cj}.

In the evolution equations for heavy quarks, Eqs.~\eqref{eq:RBTeqHQ}, we treat the phase-space distribution functions of the bulk medium $f_{j,k}(x,p)$ as external quantities through ${\cal{C}}[f_j,f_k,f_{Q}]$ and we adopt the established approximation of neglecting collisions between heavy quarks.
The HQs interact with the medium by means of only $2\rightarrow2$ elastic processes using scattering matrices calculated at Leading-Order in pQCD; nevertheless, a successful way to treat non-perturbative effects in heavy-quark scattering is given by the QPM \cite{Plumari:2011mk}.
In our approach the effective coupling $g(T)$ leads to effective vertices and a dressed massive gluon propagator for $g + HQ \to g + HQ$ and massive quark propagator for $q + HQ \to q + HQ$ scatterings.
The detail of the calculations can be found in Ref. \cite{Scardina:2017ipo}.

The hadronization process plays a crucial role in determining the final spectra, $R_{\rm AA}(p_T)$ and $v_2(p_T)$. When the temperature of the fireball goes below the quark-hadron transition temperature, $T_c=155$ MeV, we hadronize the charm quark to $D$-meson by means of a fragmentation model, see Ref.~\cite{Plumari:2017ntm} for more details.
The multiplicity is chosen by matching it with the experimental value of the $dN/dy$ for the considered centrality class. 

We perform simulations with our default initialization and simulations with modified initial conditions that allow to implement the onset of vorticity in the fireball due to the angular momentum of the system. 
The latter modelling is explained in detail in Section \ref{ic_vort}, whereas our standard initialization is as follows.

The initial conditions for the bulk in the coordinate space are given by the standard Glauber model assuming boost invariance along the longitudinal direction.
In momentum space they are given by a Boltzmann-J\"uttner distribution function up to transverse momentum $p_T^{mj}$ while at larger momenta we adopt mini-jet distributions as calculated by pQCD at NLO order in \cite{Greco:2003xt}: we take $p_T^{mj}=2.0$ GeV at RHIC and $p_T^{mj}=3.5$ GeV at LHC energy. 
The charm and anticharm quark distributions are initialized in coordinate space according to the number of binary nucleon-nucleon collisions $N_{coll}$. 
In momentum space we use charm quark production according to the Fixed Order + Next-to-Leading Log (FONLL) calculation \cite{Cacciari:2012ny} which describes the $D$-meson spectra in proton-proton collisions after fragmentation \cite{Scardina:2017ipo}.

In the last decade a lot of studies have been performed to determine the electromagnetic field generated in uRHICs \cite{Kharzeev:2007jp, Skokov:2009qp, Voronyuk:2011jd, Tuchin:2013ie, Bzdak:2011yy, Deng:2012pc, Zhong:2014sua}.
In the present work we aim to study the impact of the electromagnetic field on heavy quark dynamics.
The electromagnetic fields appearing in the kinetic equations ~\eqref{eq:RBTeqQGP} and ~\eqref{eq:RBTeqHQ} are computed as done in Ref.~\cite{Das:2016cwd}, following Ref.~\cite{Gursoy:2014aka}.
In our convention for the reference frame the nucleus A whose centre is at $x\geq0$ moves towards the positive \textit{z}-axis and the nucleus B whose centre is at $x\leq0$ moves towards the negative \textit{z}-axis. Therefore, the magnetic field ${\bm B}$ generated in non-central collisions points on average along the negative $y$ direction.
The time evolution of $B_y$ is described assuming a constant electrical conductivity $\sigma_{el}$ of the QGP in order to obtain analytic solutions of the Maxwell equations. The values of $\sigma_{el}$ employed in our calculations are typical values that can be reached in the range of temperatures explored in a heavy-ion collision, in agreement with the lattice QCD calculations
\cite{Ding:2010ga, Amato:2013naa, Brandt:2012jc}.  
Due to Faraday induction an electric field $\bf E$ along the $x$ axis is generated. 

The total magnetic field is the sum of $B_y^{+,-}$ generated by $Z$ point-like charges as ~\cite{Kharzeev:2007jp}
\begin{eqnarray}\label{totBy}
e  B_{y}(\tau,\eta_s) &=&  -Z\int_{-\frac{\pi}{2}}^{\frac{\pi}{2}} d\phi' \int_{x_{\rm in}(\phi')}^{x_{\rm out}(\phi')} d\xp' \xp' \rho_-(\xp') \nonumber \\
&\times& (eB^+_{y}(\tau,\eta,\xp,\phi) + eB^-_{y}(\tau,\eta_s,\xp,\phi))
\end{eqnarray}
In the above equation $B^+_{y}(\tau,\eta_s,\xp,\phi)$ and $B^-_{y}(\tau,\eta_s,\xp,\phi)$ are the elementary magnetic fields generated by a single charge $e$ located in the transverse plane at ${\bm\xp}=(\xp,\phi)$ and moving towards $+z$ and $-z$ respectively with speed $\beta$ related to the longitudinal space-time rapidity $\eta_s \equiv \arctan(z/t)$; $x_{\rm in}$ and $x_{\rm out}$ are the endpoints of the $\xp'$ integration regions given by 
\begin{eqnarray}\label{xpm} 
x_{\rm in/out}(\phi') = \mp \frac{b}{2}\cos(\phi') + \sqrt{R^2-\frac{b^2}{4}\sin^2(\phi')}\, ,
\end{eqnarray}
where $R$ is the radius of the nucleus and $b$ is the impact parameter of the collision.
The elementary electromagnetic fields are obtained by solving the Maxwell equations and can be written as
\begin{eqnarray}\label{Bsingle}
e B^+_y(\tau,\eta_s,\xp,\phi) &=& \alpha\sinh(y)(\xp \cos\phi -  x'_\perp \cos\phi') \nonumber\\
&\times& \frac{\sigma_{el}\,\frac{|\sinh(y)|}{2} \xi^{\frac12} +1}{\xi^{\frac32}} e^A\, ,
 \end{eqnarray}
 where $\alpha=e^2/(4\pi)$ is the electromagnetic coupling and $y\equiv \arctan(\beta)$ is the beam rapidity of the $+$ mover. In the above equation $A$ and $\xi$ are given by 
\begin{eqnarray} \label{ADelta}
A &\equiv&  {\frac{\sigma_{el}}{2}\left(\tau\sinh(y)\sinh(y-\eta_s)-|\sinh(y)| \xi^{\frac12}\right)}\,, \nonumber \\  
 \xi &\equiv& \tau^2\sinh^2(y-\eta_s) + \xp^2+\xp^{'2}\nn - 2\xp\xp'\cos(\phi-\phi')\,.
\end{eqnarray}
In a similar way an electric field produced by the charges moving along the $z$ direction can be obtained as 
\begin{equation}\label{Esingle}
e E^+_x(\tau,\eta_s,\xp,\phi) = \,  e B^+_y(\tau,\eta_s,\xp,\phi) \coth(y-\eta_s)\,;
\end{equation}
this field has to be convoluted with the transverse charge distribution $\rho_\pm(\xp)$ as for the magnetic field.

\begin{figure}[!hbt]
\centering
\includegraphics[width=0.95\columnwidth]{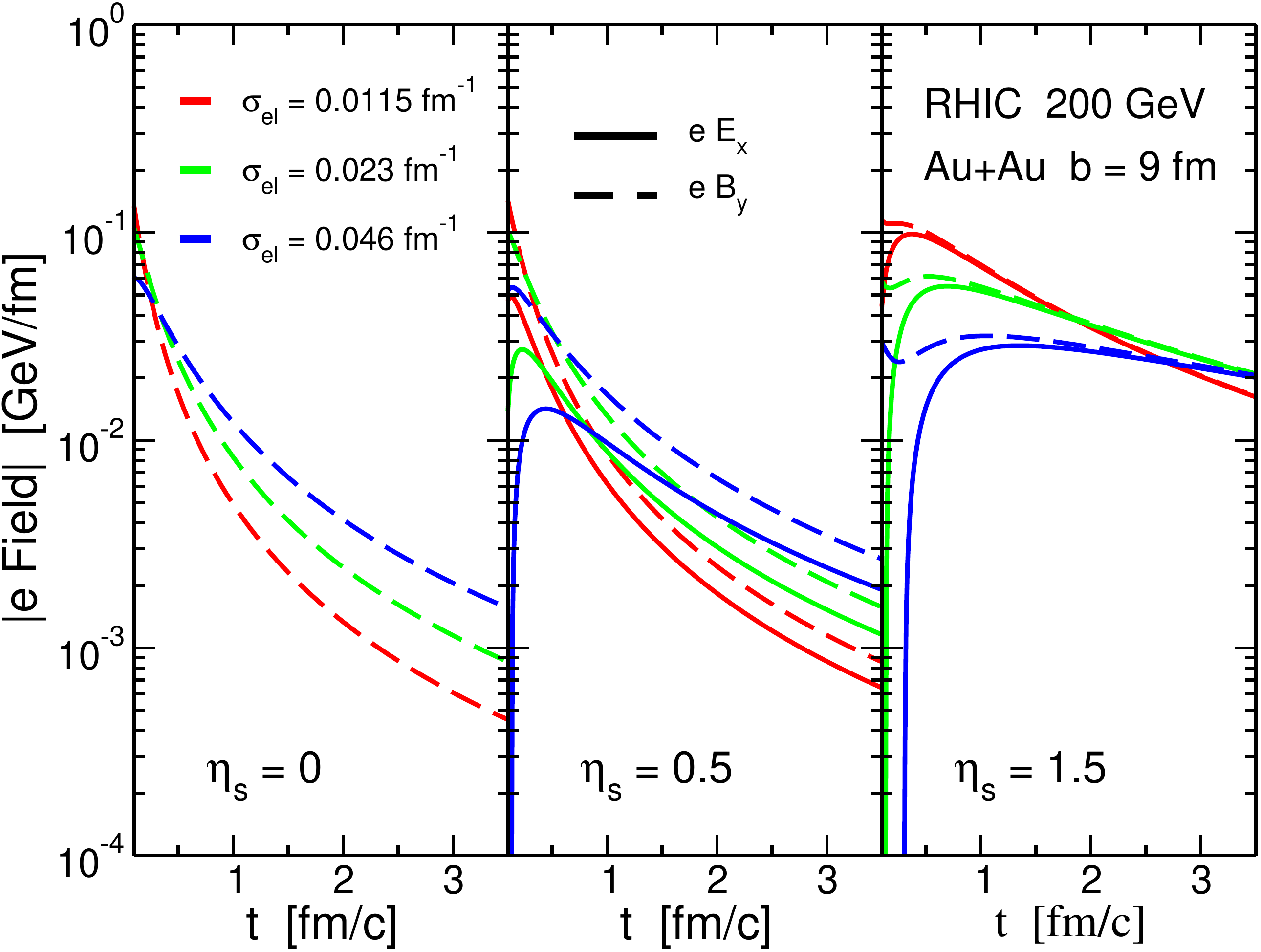}
\\[\baselineskip]
\includegraphics[width=0.95\columnwidth]{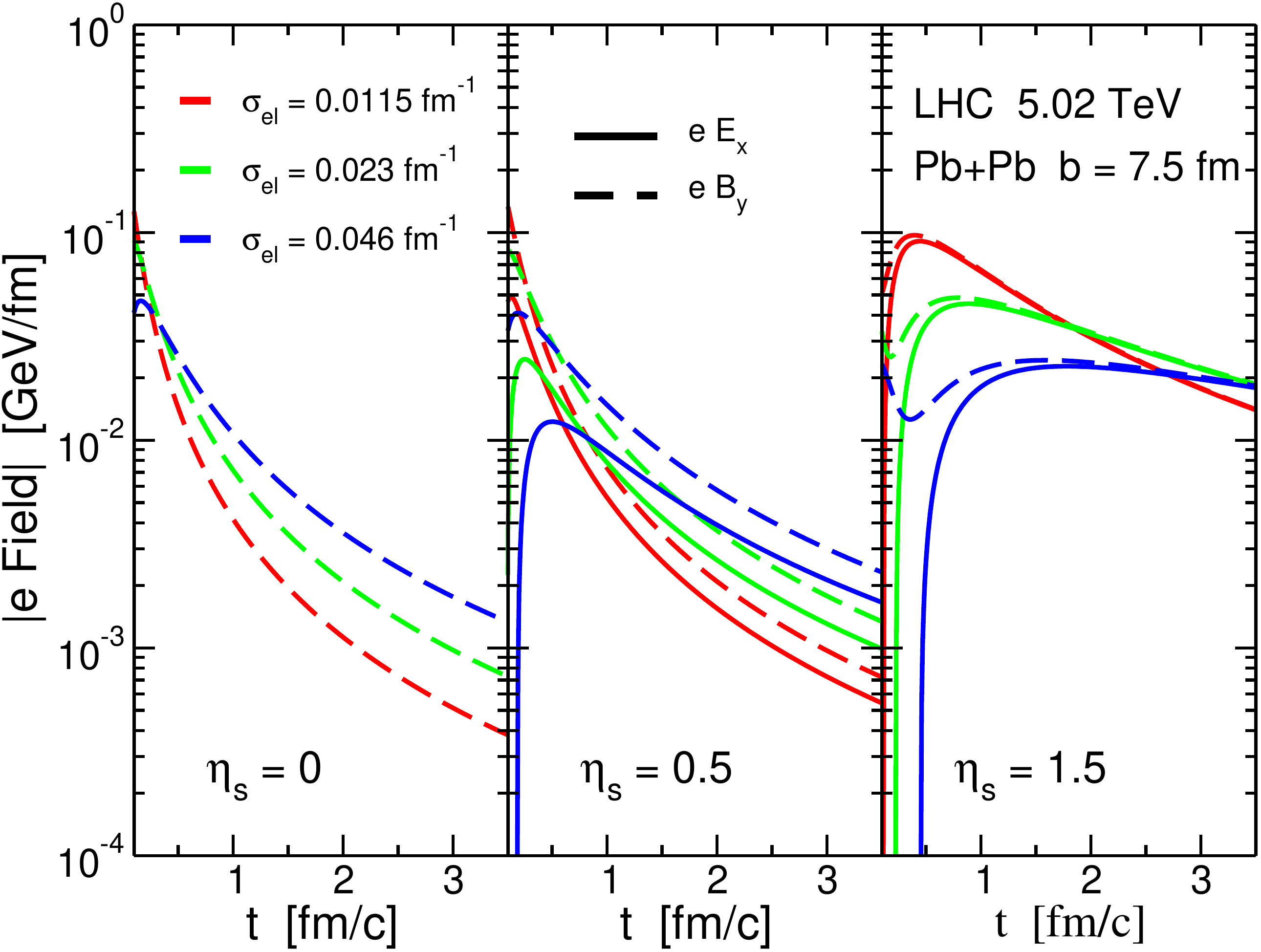}
\caption{Time dependence of $\vert e E_x\vert$ (solid curves) and $\vert e B_y\vert$ (dashed curves) in Au+Au collisions at $\sqrt{\sigma_{NN}}=200$ GeV with $b=9$ fm (top panel) and Pb+Pb collisions at $\sqrt{\sigma_{NN}}=5.02$ TeV with $b=7.5$ fm (bottom panel). The computation is done at $\xp=0$ and different points in space-time rapidity $\eta_s$. The different colors correspond to different values of the electric conductivity $\sigma_{el}$.}
\label{fig:EMF_rhicb9}
\end{figure}

In Fig.~\ref{fig:EMF_rhicb9} we show the time evolution of $E_x$ and $B_y$ in Au+Au collisions at $\sqrt{\sigma_{NN}}=200$ GeV with impact parameter $b=9$ fm (top panel) and Pb+Pb collisions at $\sqrt{\sigma_{NN}}=5.02$ TeV with $b=7.5$ fm (bottom panel). The curves are calculated at ${\bm\xp}=0$ for three values of the space-time rapidity: $\eta_s=0$ (left panel), $\eta_s=0.5$ (middle panel) and $\eta_s=1.5$ (right panel). Red, green and blue lines correspond respectively to $\sigma_{el}=0.0115,\,0.023,\,0.046$ fm$^{-1}$ which correspond to values of the conductivity as computed in lattice QCD \citep{Amato:2013naa, Buividovich:2010tn,Puglisi:2014pda}. 
From the left panel we see that at $\eta_s=0$ the electric field vanishes due to symmetry. However, at forward and backward rapidity $E_x$ become huge and comparable to $B_y$, as we can see from the middle and right panels in which the results at positive rapidities are shown; for symmetry reasons, at the corresponding points in the negative $\eta_s$ plane $B_y$ is the same and $E_x$ has opposite sign: $B_y(\eta_s)=B_y(\eta_s)$ and $E_x(-\eta_s)=E_x(\eta_s)$.
At $\eta_s=0$ the magnetic field $\vert eB_y\vert$ reduces initially by about one order of magnitude in 1 fm$/c$ and the decrease is higher for smaller electric conductivity. At larger $\eta_s$ the decrease of $\vert eB_y\vert$ becomes smaller, especially at $\eta_s=1.5$.
It is important to notice that the electromagnetic field at RHIC and LHC is quite similar and has a maximum value that is about
0.1 GeV/fm, however in vacuum the maximum at RHIC is some factor larger while at LHC is more than 50 times larger.

Aside from $E_x$ and $B_y$, the other components of the electromagnetic field averaged over many initial conditions vanish or are very small.
However, in event-by-event collisions the large fluctuations in the positions of the protons inside the two colliding nuclei can generate non-zero values of the other components of the electromagnetic field with magnitude comparable but generally smaller than $B_y$ and $E_x$ \cite{Bzdak:2011yy,Deng:2012pc}.

Furthermore, for a more complete calculation we should also include the electromagnetic field generated by the participant protons; however, it has been shown in \cite{Voronyuk:2011jd, Gursoy:2014aka} that it is subdominant in magnitude with respect to that produced by spectators, especially in the early stage that is crucial for the formation of a directed flow of heavy mesons that is the focus of this work.

Within our transport approach we are able to simultaneously describe the $R_{AA}(p_T)$ and the $v_2(p_T)$ of $D$ mesons both at RHIC and LHC energies \cite{Scardina:2017ipo}. The inclusion of the electromagnetic field does not affect visibly those quantities, while it produces a sizeable effect on the rapidity dependence of the directed flow $v_1$ of $\Dz$ and $\Dzb$ as we will see in the following sections.

\section{Set up of the initial condition}
\label{ic_vort}

In this subsection we discuss how we set up the initial vortical structure of the fireball produced in uRHICs.

In a non-frontal collision the system constituted by the two incoming nuclei possesses an angular momentum which depends mainly by the energy, the size and the impact parameter of the collision. Only a fraction of this angular momentum is transferred to the plasma created after the collision, since most of it is carried away by the spectator nucleons. The angular momentum retained by the QGP manifest itself as a nonzero vorticity of the system \cite{Becattini:2007sr, Csernai:2013bqa, Becattini:2015ska, Deng:2016gyh, Jiang:2016woz}.
Even if the total angular momentum $J$ of the plasma, which quantifies its global rotation, could be precisely determined, several phase-space configurations of the plasma would correspond to that value of $J$. Moreover, it is not well known so far how this vortical structure is produced in the plasma.

In hydrodynamical simulations $J$ has been introduced as an asymmetric initial energy density distribution respect to the reflection $\eta_s\rightarrow -\eta_s$ \cite{Bozek:2010bi, Becattini:2015ska}. In those initial conditions the initial flow velocity Bjorken components are zero, but the longitudinal asymmetry of the initial energy density profile generates fluid vorticity.
At variance with the common use of hydrodynamic models, many simulations based on relativistic kinetic theory generate automatically a vorticity in the QGP since positions and momenta of partons are determined by the primary collision of the two incoming nuclei \cite{Deng:2016gyh, Jiang:2016woz, Teryaev:2015gxa, Kolomeitsev:2018svb}.
Hence, the initial longitudinal velocity distribution of the fireball is asymmetric with respect to the reflection $\eta_s\rightarrow -\eta_s$ and the induced shear flow produces a nonzero local vorticity.

In this paper we extend our model in order to take into account the nonzero angular momentum that is transferred to the QGP by the colliding nuclear system. Comparing it to our default initialization that does not include any effect of the angular momentum of the collision we are able to disentangle the effect of vorticity on dynamical evolution and final observables of charged particles and heavy mesons.
The default type of initial conditions, introduced in the previous section, is given by a full longitudinally boost invariant distribution, i.e. a uniform pseudorapidity distribution in the range $\eta_s\in[-2,2]$ for RHIC and $\eta_s\in[-3,3]$ for LHC and a standard transverse distribution based on the Glauber model at the initial time $t_0$ of the simulation.
The second initialization is an extension of the previous one in order to include the angular momentum generated in noncentral collisions, inspired by the modelling adopted within the hydrodynamic framework in Ref.~\cite{Bozek:2010bi, Becattini:2015ska}.

Within our convention, the nucleus A whose centre is at $x\geq0$ moves towards the positive \textit{z}-axis respectively and the nucleus B whose centre is at $x\leq0$ moves towards the negative \textit{z}-axis.
We implement a vortical structure in the QGP modifying the longitudinally boost invariant Bjorken picture with an initial density profile asymmetric with respect to the reflection $\eta_s\rightarrow -\eta_s$:
\begin{equation}\label{eq:ini_dens}
\rho({\bm\xp},\eta_s)=\rho_0\dfrac{W({\bm\xp},\eta_s)}{W(0,0)} 
H(\eta_s)
\end{equation}
where ${\bm\xp}$ is the transverse coordinate, $\rho_0=\rho(0,0)$ is the density at the centre of the fireball and $W$ is the wounded nucleon weight function given by
\begin{equation}\label{eq:wounded}
W({\bm\xp},\eta_s)=2\left( N_A({\bm\xp})f_-(\eta_s)+N_B({\bm\xp})f_+(\eta_s)\right).
\end{equation}
$N_A$ and $N_B$ are the number of participant nucleons of the nuclei A and B respectively; they are determined from the Glauber model in the following way:
\begin{eqnarray}
N_A({\bm\xp})=T_A({\bm\xp})\left(1-\mathrm{e}^{-\sigma_{NN}^{in}T_B({\bm\xp}-{\bm b})} \right),\\
N_B({\bm\xp})=T_B({\bm\xp}-{\bm b})\left(1-\mathrm{e}^{-\sigma_{NN}^{in}T_B({\bm\xp})} \right), 
\end{eqnarray}
where $\sigma_{NN}^{in}$ is the inelastic nucleon-nucleon cross section taken to be 4.2 fm$^2$ at RHIC energy $\sqrt{\sigma_{NN}}=200$ GeV and 6.8 fm$^2$ at LHC energies $\sqrt{\sigma_{NN}}=5.02$ TeV; $T_A$ and $T_B$ are the thickness functions of the nuclei A and B, respectively, defined as in standard Glauber model. 
In Eq.~\eqref{eq:ini_dens} the function
\begin{equation}\label{eq:Hfunct}
H(\eta_s)=\exp \left[ -\dfrac{(|\eta_s|-\eta_{s0})^2}{2\sigma_\eta^2}\theta(|\eta_s|-\eta_{s0})\right]
\end{equation}
determines in the initial density distribution a central plateau of length $2\eta_{s0}$ and gaussian tails at larger rapidity of width $\sigma_\eta$. These two parameters are fixed to better reproduce the experimental pseudorapidity density of charged particles $dN_{ch}/d\eta$. A complete description of the experimental tails in the fragmentation regions cannot be achieved within our model which involves only the partonic phase, however our initial condition allow us to reproduce fairly well the $dN_{ch}/d\eta$ in the central rapidity region $|y| \lesssim 1.5$ which is of interest for our work.\\
The functions $f_\pm(\eta_s)$ in Eq.~\eqref{eq:wounded}, representing the emission contributions respectively from forward-going and backward-going participant nucleons, are given by
\begin{equation}\label{eq:f+f-}
f_+(\eta_s)=f_-(-\eta_s)=
\begin{cases}
\quad 0                        & \quad \eta_s<-\eta_m        \\
\dfrac{\eta_s+\eta_m}{2\eta_m} & -\eta_m\leq\eta_s\leq\eta_m \\
\quad 1                        & \quad \eta_s>\eta_m
\end{cases}
\end{equation}
The parameter $\eta_m$ in the functions $f_\pm(\eta_s)$ determines an asymmetry in the contribution to the local participant density from the forward and backward-going nuclei and this leads to a tilted fireball in the reaction plane \cite{Bozek:2010bi} which owns a huge angular momentum reflecting that transferred by the two initial nuclei during the collision. This initial spatial asymmetry of the fireball along the longitudinal direction is translated, by means of the collective evolution and expansion, to a final momentum distribution of the particles with a nonzero directed flow $v_1$ that is odd in pseudorapidity $\eta$.
The angular momentum ${\bm J}$ of the plasma is not a directly measurable quantity and, moreover, the same value of ${\bm J}$ can be obtained by means of different realizations of the initial conditions. It has been shown within the hydrodynamical framework \cite{Bozek:2010bi, Becattini:2015ska} that the tilted initial source parametrized by means of Eq.~\eqref{eq:f+f-} allow to correctly reproduce the $\eta$ dependence of the directed flow of charged particle in peripheral relativistic collisions, once a suitable value of the parameter $\eta_m$ is chosen.
As every initialization that generates a vorticity in the system, this kind of initial conditions breaks the boost invariance on the longitudinal direction since the distribution of particles according to Eqs.~\eqref{eq:ini_dens}--\eqref{eq:f+f-} is not uniform in space-time rapidity $\eta_s$ even in the central range.

\begin{figure}[!hbt]
\centering
\includegraphics[width=0.95\columnwidth]{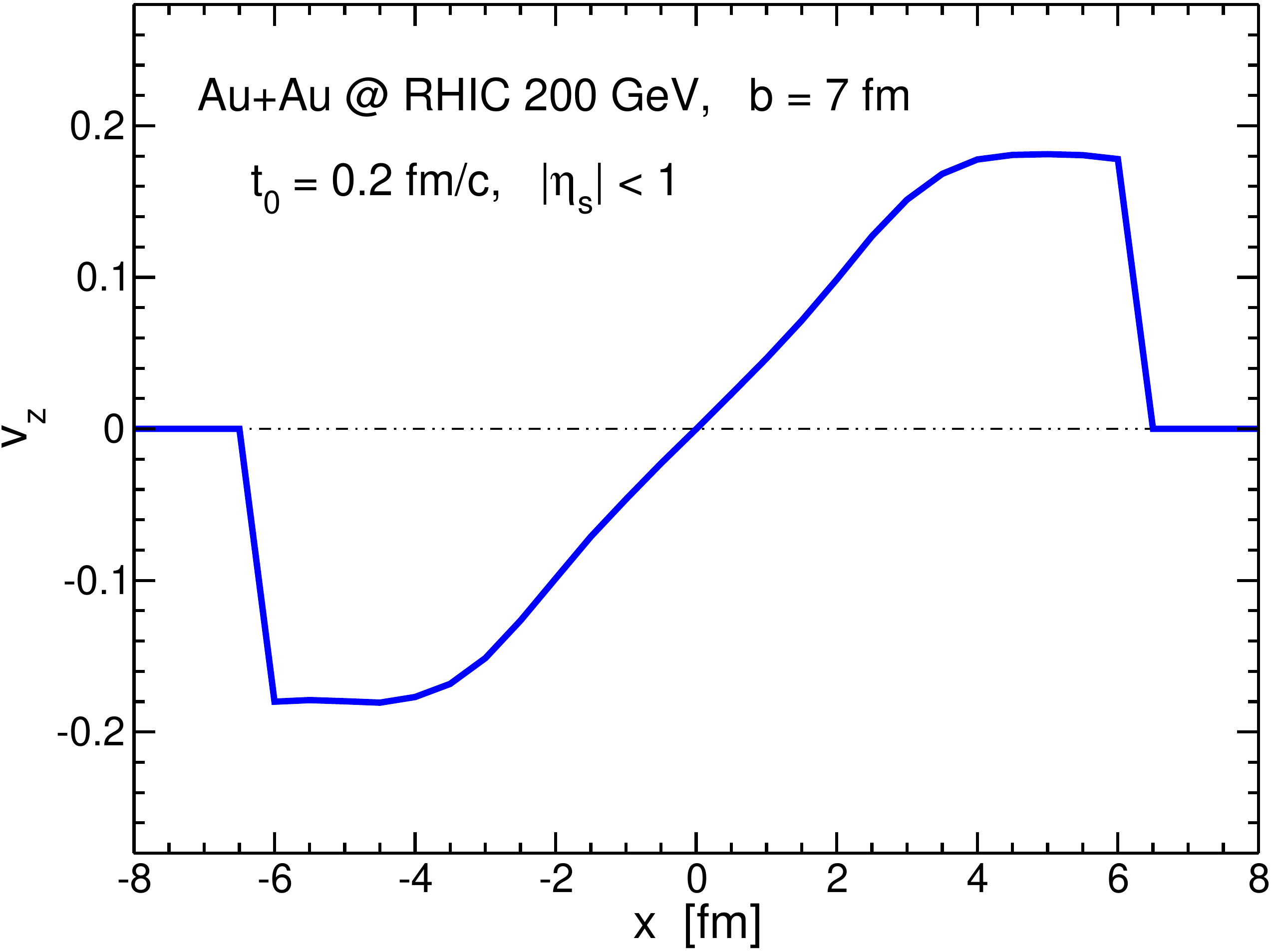}
\caption{Longitudinal velocity profile along the impact parameter axis at the initial time of the simulation $t_0=0.2$ fm$/c$ for Au+Au collisions at top RHIC energy $\sqrt{\sigma_{NN}}=200$ GeV with impact parameter $b=7$ fm.}
\label{fig:vz_x}
\end{figure}

\begin{figure}[!hbt]
\centering
\includegraphics[trim={0 5 0 0},clip,width=0.9\columnwidth]{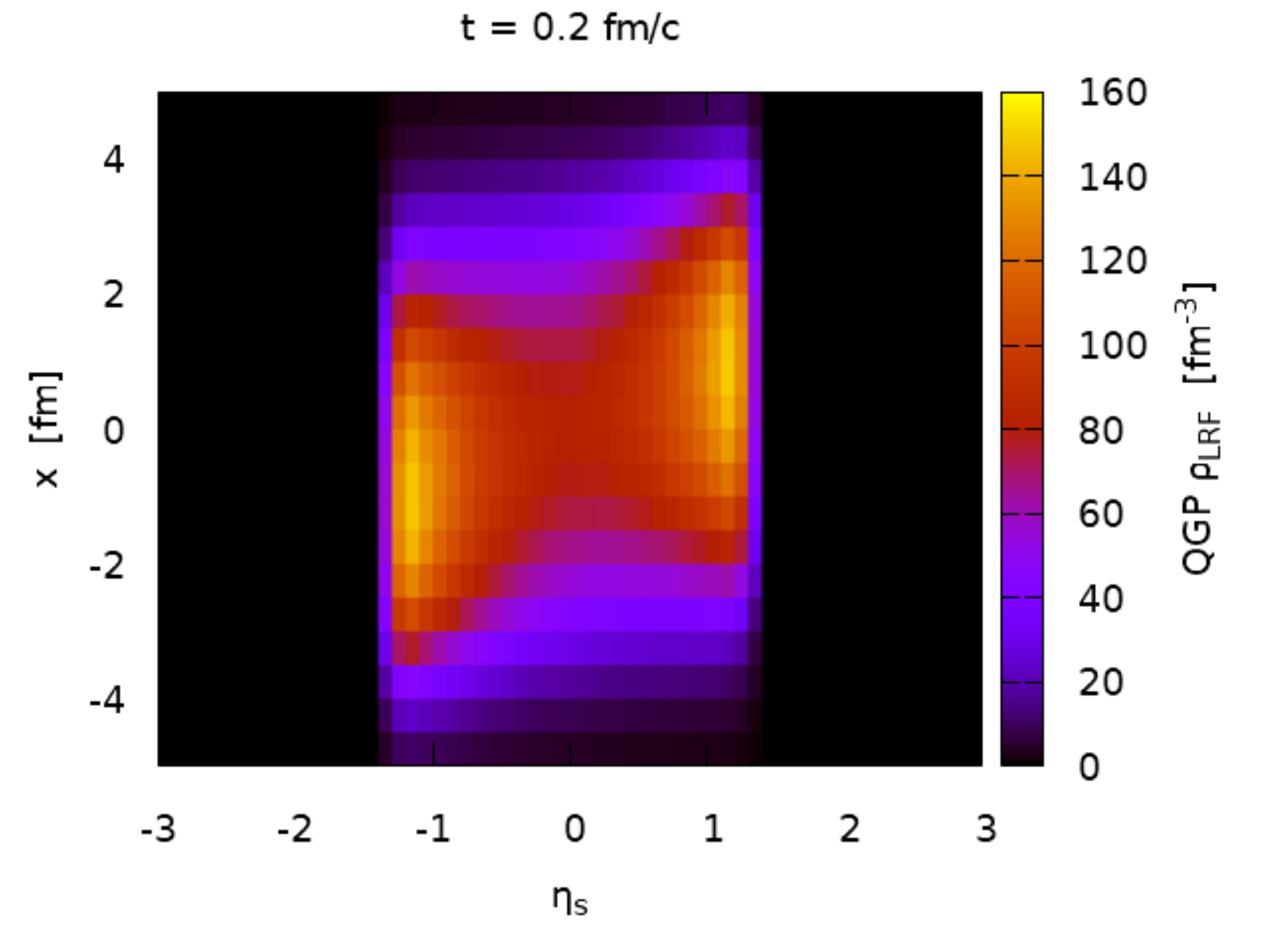}
\\[\baselineskip]
\includegraphics[trim={0 5 0 0},clip,width=0.9\columnwidth]{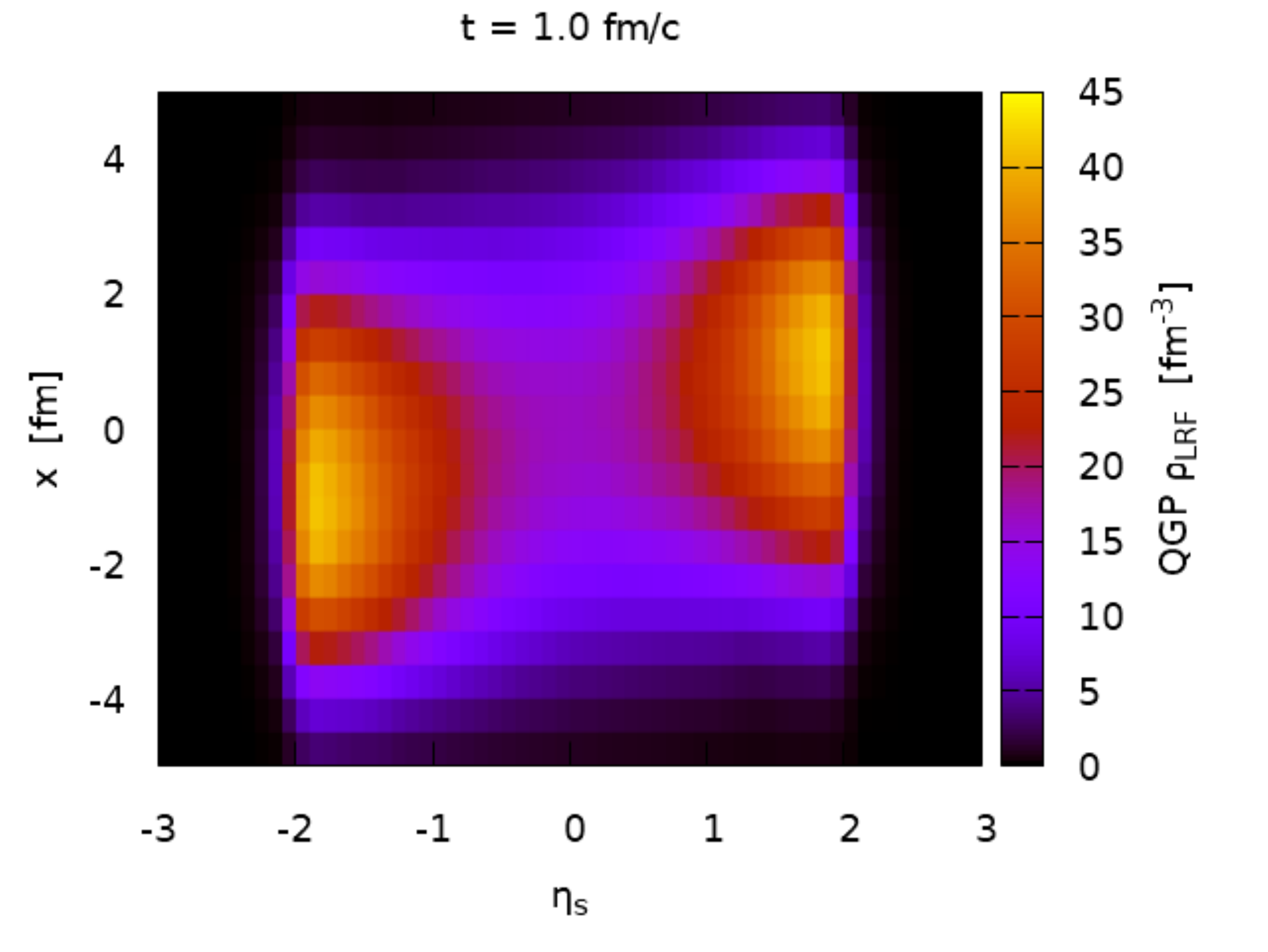}
\\[\baselineskip]
\includegraphics[trim={0 5 0 0},clip,width=0.9\columnwidth]{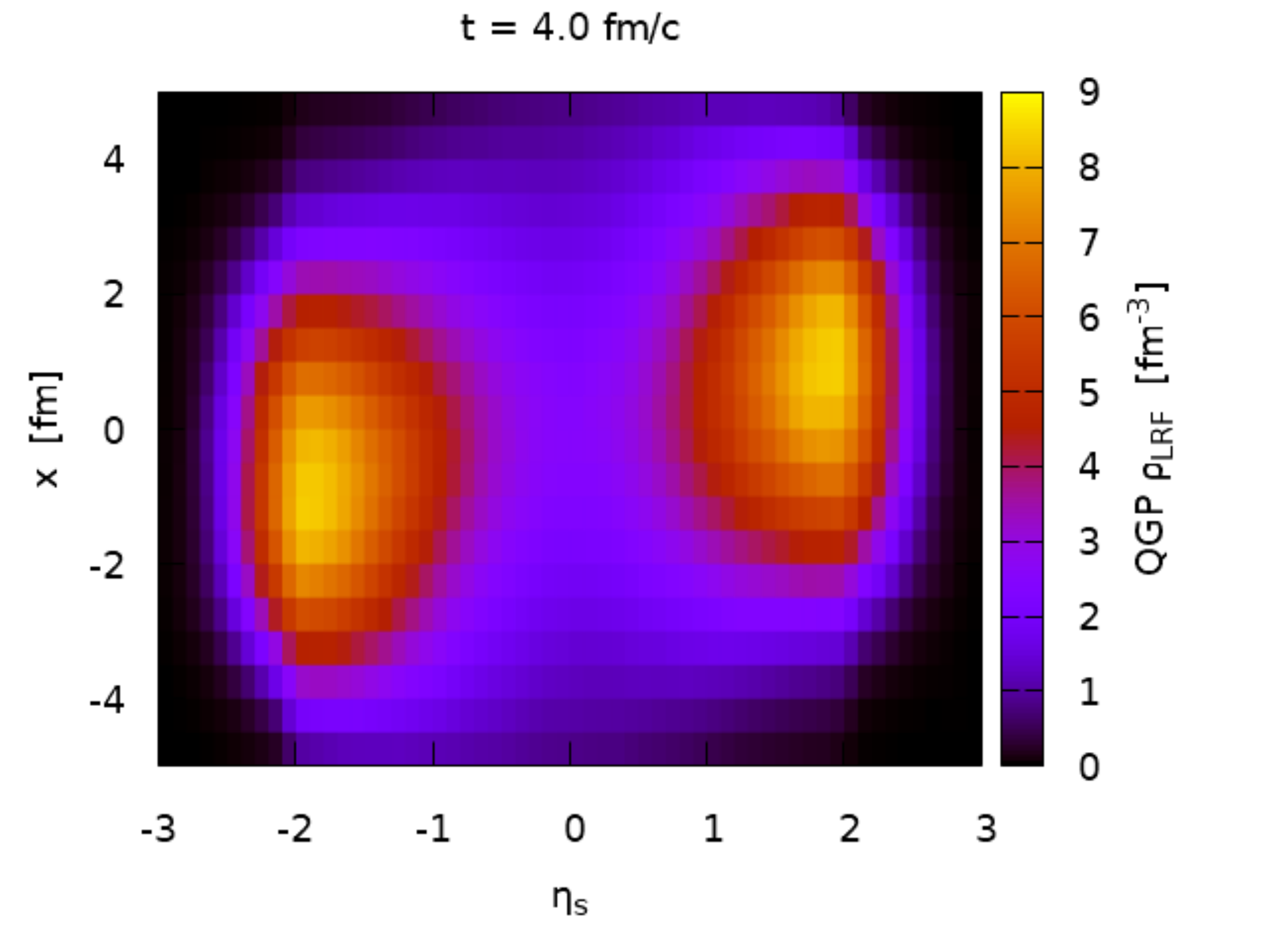}
\caption{Profiles of the proper density in the reaction plane of the quark-gluon plasma at the initial time of the simulation $t_0=0.2$ fm$/c$ (top panel) and at the later times $t=1$ fm$/c$ (middle panel) and $t=4$ fm$/c$ (bottom panel).}
\label{fig:tilted}
\end{figure}

\begin{figure}[!hbt]
\centering
\includegraphics[width=0.95\columnwidth]{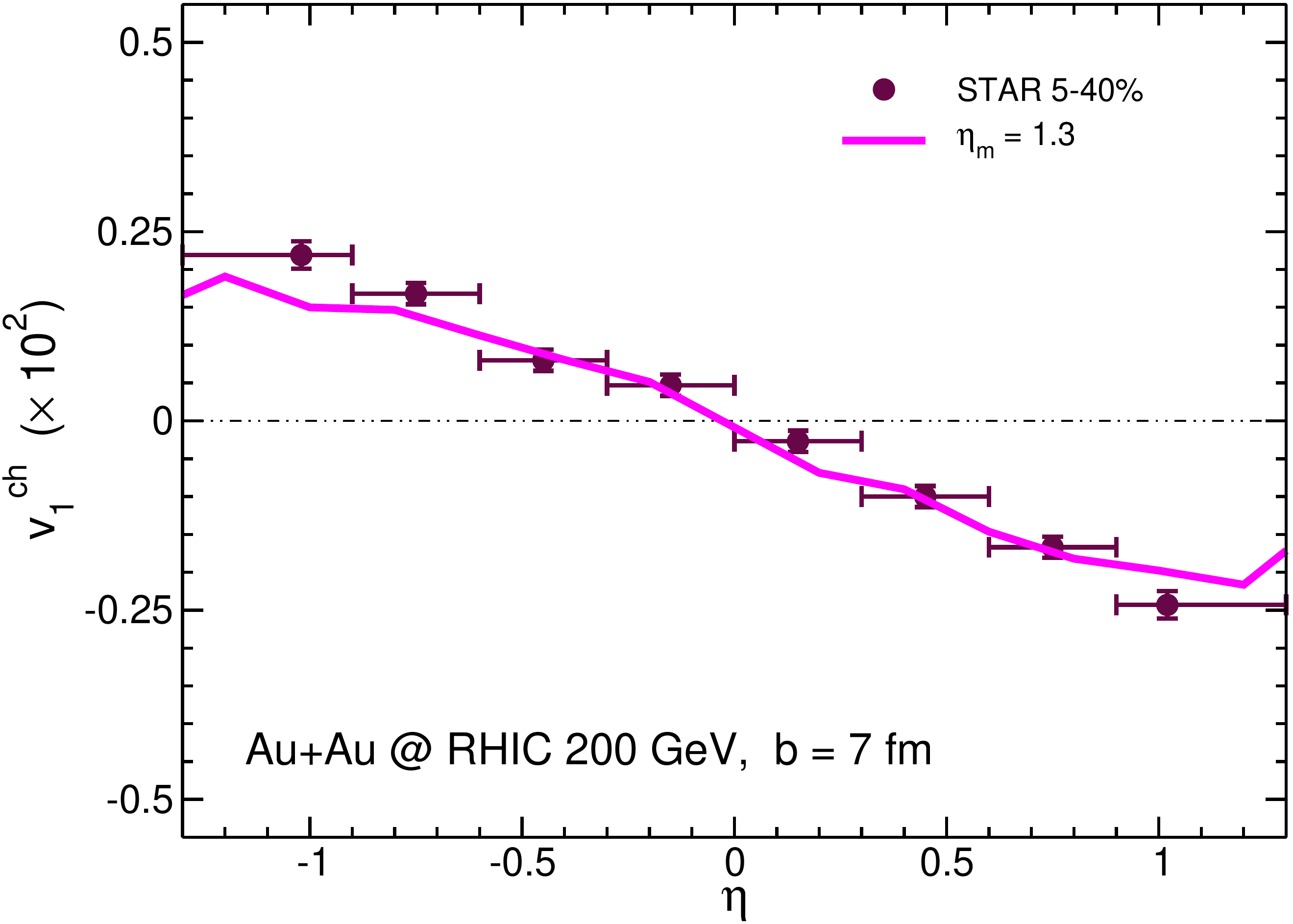}
\caption{Directed flow of charged particles versus pseudorapidity for collisions at top RHIC energy $\sqrt{\sigma_{NN}}=200$ GeV with impact parameter $b=7$ fm (magenta line) in comparison to the experimental data from STAR Collaboration in the centrality bin 5-40\% \cite{Abelev:2008jga} (maroon circles).}
\label{fig:v1ch_rhicb7}
\end{figure}

For Au+Au collisions at RHIC energy of $\sqrt{\sigma_{NN}}=200$ GeV we consider an initial time $t_0=0.2$ fm$/c$ of the simulation and a maximum initial temperature at the center of the fireball $T_0=410 $ MeV; for 
the impact parameter $b=7$ fm we use the values $\eta_{s0}=1.0$, $\sigma_\eta=1.3$ and $\eta_m=1.3$ for the parameters in Eqs.~\eqref{eq:Hfunct}--\eqref{eq:f+f-}.
\\
In Fig.~\ref{fig:vz_x} we show the longitudinal velocity profile along the impact parameter axis of the bulk medium constituted by gluons and light quarks in $\vert\eta_s\vert<1$ at the initial time of the simulation $t_0=0.2$ fm$/c$.
While with our standard initialization $v_z(\eta_s)=-v_z(-\eta_s)$ and the longitudinal velocity profile of a symmetric $\eta_s$-slice vanishes all along the impact parameter axis, the initial conditions previously explained generate a shear flow of the different $x$-layers of the fluid along the beam direction.
\\
The corresponding initial distribution on the $\eta_s-x$ plane of the proper density of the bulk matter at the initial time of the simulation $t_0=0.2$ fm$/c$ is depicted on the top panel of Fig.~\ref{fig:tilted}; it is evident how the asymmetry in the forward-backward hemispheres induced by the emission functions $f_\pm(\eta_s)$ leads to a tilt of the fireball in the reaction plane. The middle and bottom panels show the evolution of the proper density profile at two later times, $t=2$ fm$/c$ and $t=4$ fm$/c$ respectively.

The information on how this spatial anisotropy is transferred to the momentum space is encoded on the rapidity-odd directed flow of charged particle $v_1^{ch}$, which we show as a function of pseudorapidity in Fig.~\ref{fig:v1ch_rhicb7}; our result (magenta line) fairly matches the STAR experimental data for 5--40\% central collisions \cite{Abelev:2008jga} (maroon circles) in the range $|\eta|<1$.
It is important to note that the bulk QGP medium does not rotate like a rigid body but the angular momentum of the system constituted by the two initial nuclei is transferred to the plasma as a longitudinal shear flow, differently with respect to what happens in heavy-ion collisions at lower energy where a significant space rotation of the participant region or the strong bouncing-off in the transverse direction determine a quite sizeable transverse flow  \cite{Reisdorf:1997fx, Andronic:2006ra}.

As mentioned in the previous section, the charm and anticharm quark distributions are initialized in coordinate space according to hard binary scatterings between nucleons of the initial colliding nuclei. Hence, their production is symmetric with respect to the reflection $\eta_s \rightarrow -\eta_s$ and does not follow the profile introduced in Eqs.~\eqref{eq:Hfunct}--\eqref{eq:f+f-} for characterizing the distribution of gluons and light quarks.
In Ref.~\cite{Chatterjee:2017ahy} this shift between the spatial profiles of bulk matter and binary collisions was introduced. 
It implies that at $\eta_s \neq 0$ the area of heavy quark production points in the transverse plane is shifted with respect to the transverse section of the fireball, and the shift increases at higher absolute values of the space-time rapidity. The asymmetrically-distributed bulk matter drags the heavy quarks, whose directed flow results to be amplified and about one order of magnitude larger than that of charged particles.
\\
For a more sophisticated calculation we should include, from one hand, a dependence on $N_{coll}$ of the wounded nucleon weight function in Eq.~\eqref{eq:wounded} for the energy deposition of the bulk matter in order to take into account the contribution from binary collisions and, from the other hand, the heavy quark production profile may present a mild asymmetry due to fluctuations in nucleon positions in the colliding nuclei; for the sake of simplicity, we neglect both effects. These are however expected to be subdominant with respect to the main mechanism generating the large $v_1$ of heavy quarks.

\section{Vorticity of the QGP}
\label{vortQGP}

Due to the initial condition explained in the previous section the fireball owns a nonzero angular momentum and a strong vortical flow.
In this section we focus on Au+Au collisions at top RHIC energy $\sqrt{\sigma_{NN}}=200$ GeV with impact parameter $b=7$ fm; in all plots we show quantities averaged over about 1000 events.

The total orbital angular momentum ${\bm J}$ of the QGP can be computed summing the contributions ${\bm J}_i$ of all particles:
\begin{equation}
{\bm J}=\sum_i{\bm J}_i=\sum_i{\bm r}_i\times{\bm p}_i,
\end{equation}
where ${\bm r}_i$ is the position vector of the particle respect to the centre of the overlap region and ${\bm p}_i$ is its momentum.
We find that ${\bm J}$ is mainly directed perpendicularly to the reaction plane; due to our convention of the reference frame its dominant component $J_y$ lies along the negative $y$ axis.
Regarding the event-averaged values, for RHIC collisions at $\sqrt{\sigma_{NN}}=200$ GeV with $b=7$ fm we find $J_y\approx-10400$ ($\hslash$ units) while the other components are four order of magnitude smaller, in agreement with the picture expected and discussed in previous works \cite{Becattini:2007sr, Csernai:2013bqa, Becattini:2015ska, Deng:2016gyh, Jiang:2016woz}.
However, in each event the x and z components can reach values of few hundred $\hslash$.
For what concern the time dependence, we have checked that the total angular momentum stay almost constant during the whole evolution, being conserved within 0.2\,\%.
Our result for $J_y$ is in agreement with calculations based on the HIJING model \cite{Deng:2016gyh}.
However, a strict comparison of the magnitude of the QGP angular momentum computed through different approaches is not very meaningful; indeed, the precise value of $J$ depends on several factors, first of all on the size of the fireball along the longitudinal direction, and a small difference in that length could produce a very different amount of angular momentum.

\begin{figure}
\centering
\includegraphics[trim={5 5 0 0},clip,width=0.9\columnwidth]{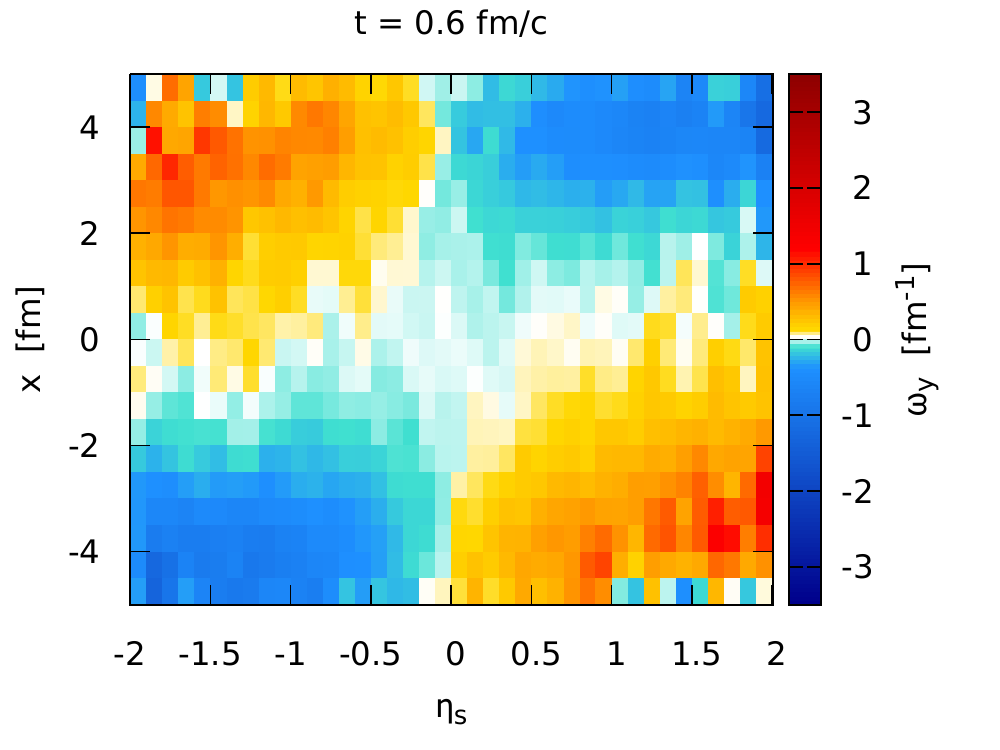}
\\[\baselineskip]
\includegraphics[trim={5 5 0 0},clip,width=0.9\columnwidth]{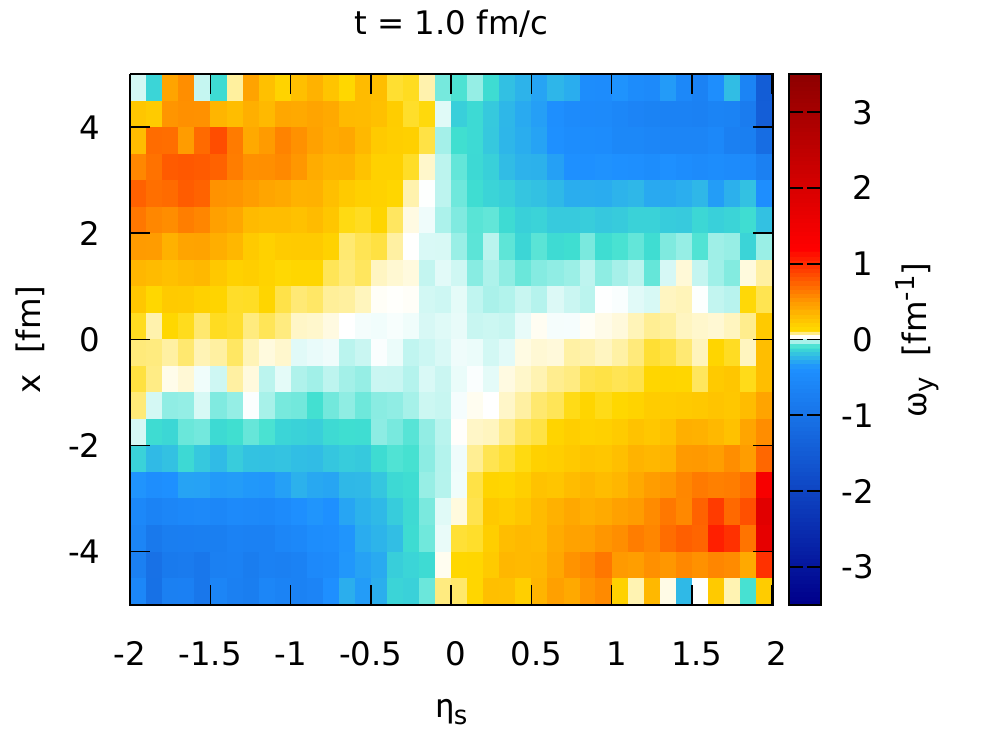}
\\[\baselineskip]
\includegraphics[trim={5 5 0 0},clip,width=0.9\columnwidth]{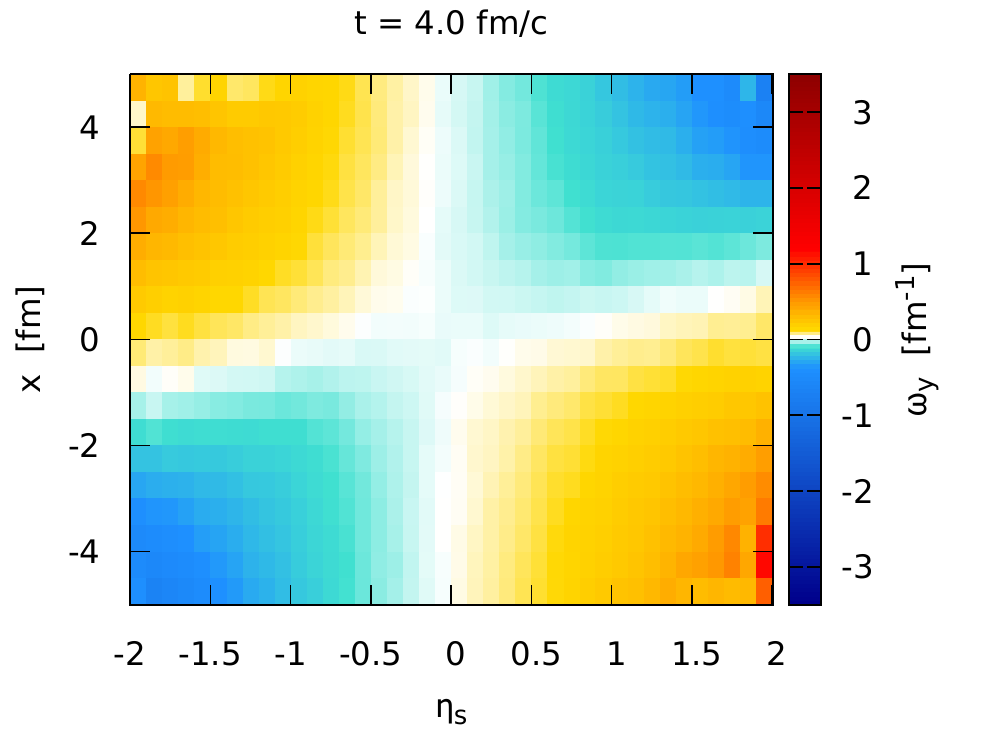}
\caption{Event-averaged non-relativistic vorticity $\omega_y$ in the reaction plane $\eta_s-x$ at the different times $t=0.6,\,1,\,4$ fm/$c$ for Au+Au collisions at top RHIC energy $\sqrt{\sigma_{NN}}=200$ GeV with impact parameter $b=7$ fm.}
\label{fig:vort_etax}
\end{figure}

A more suitable quantity for comparing different models is the local vorticity distribution. 
The classical vorticity is defined as the curl of the velocity field ${\bm v}$:
\begin{equation}
{\bm \omega}=\nabla\times{\bm v},
\end{equation}
being the velocity of a fluid cell identified with the velocity of the particle flow inside the cell and hence computed as ${\bm v}=(\sum_i{\bm p}_i)/(\sum_i E_i)$, where the sum runs over all particles in a given cell and ${\bm p}_i$ and $E_i$ are momentum and energy of the $i$ particle.\\
Since we are dealing with a relativistic system, it would be more appropriate to go beyond the non-relativistic computation, where several vorticities can be defined \cite{Becattini:2015ska}.
However, for the goals of this paper we do not need to use the values of the vorticity and the main aim of this section is to show that within our model we obtain a realistic magnitude and time behaviour of the vortical flow in the fireball in agreements with the ongoing research on the $\Lambda$ polarization.

In Fig.~\ref{fig:vort_etax} we plot the distribution of the component of the classical vorticity perpendicular to the reaction plane $\omega_y$ on the $\eta_s-x$ plane for $|y|<0.25$ fm at the times $t=0.6,\,1,\,4$ fm/$c$.
In agreement with other works \cite{Becattini:2015ska, Jiang:2016woz}, $\omega_y$ shows a quadrupole pattern with nearly vanishing values along the axis $x=0$ and $\eta_s=0$ and increasing absolute values for increasing $x$ and $\eta_s$.
Within our convention for the reference frame (nucleus with centre at positive $x$ moving towards positive $z$) $\omega_y$ is negative/positive in the regions where $x$ and $\eta_s$ have the same/opposite sign.

\begin{figure}[!hbt]
\centering
\includegraphics[width=0.95\columnwidth]{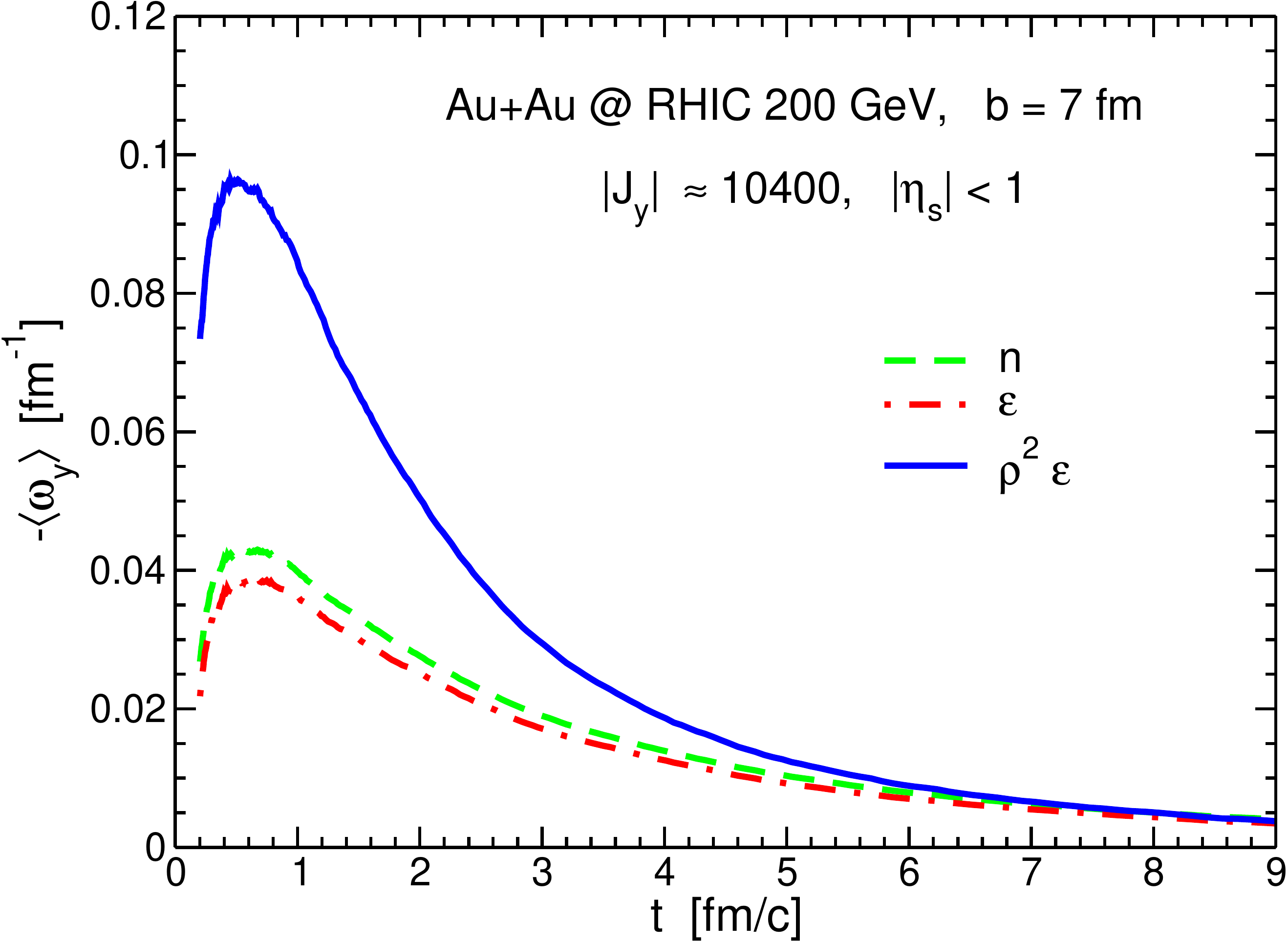}
\caption{Time evolution of the space-averaged non-relativistic vorticity $\langle\omega_y\rangle$ for Au+Au collisions at top RHIC energy $\sqrt{\sigma_{NN}}=200$ GeV with impact parameter $b=7$ fm; the different lines correspond to different weighting functions in Eq.~\eqref{eq:weight_vort}: particle density $n$, energy density $\varepsilon$ and moment-of-inertia density ${\cal I}=\rho^2\varepsilon$.}
\label{fig:vort_t}
\end{figure}

However, as pointed out in Ref.~\cite{Jiang:2016woz}, a locally nonzero vorticity is not directly related to a global angular momentum and these vortical patterns could be mainly due to the radial flow of the fireball; the global rotational motion can be pinpointed by performing an average over the fireball in order to cancel out the radial flow contributions to the local vorticity and obtain a quantity that better reflects the strength of the vorticity acting on the entire interaction area \cite{Jiang:2016woz, Deng:2016gyh}.
The space-averaged vorticity is defined by
\begin{equation}\label{eq:weight_vort}
\langle\omega\rangle=\dfrac{\int d^2{\bm\xp}\int d\eta_s\,g({\bm\xp},\eta_s)\,\omega({\bm\xp},\eta_s)}{\int d^2{\bm\xp}\int d\eta_s\,g({\bm\xp},\eta_s)},
\end{equation}
where $g({\bm\xp},\eta_s)$ is a weighting function.
In Fig.~\ref{fig:vort_t} we show the time evolution of the dominant component of the classical vorticity $\omega_y$ averaged according to Eq.~\eqref{eq:weight_vort} over the whole transverse plane and space-time rapidities $|\eta_s|<1$; we consider three weighting functions: the particle density $n$ (dashed green line), the energy density $\varepsilon$ (dot-dashed red line) and a sort of moment-of-inertia density ${\cal I}=\rho^2\varepsilon$ (solid blue line), where $\rho$ is the distance of the cell from the $y$ axis.
At early times $\vert\langle\omega_y\rangle\vert$ increases and reaches a peak at about 0.5 fm/$c$; then it begins to decrease with time approaching zero at the late stage of the evolution.
We can compare our solid blue line with the same quantity computed in Ref.~\cite{Jiang:2016woz} with the AMPT model: in both cases $\vert\langle\omega_y\rangle\vert$ weighted with ${\cal I}$ evolves after 1 fm$/c$ roughly as $\sim\exp(-t/2)$ but the peak value in our model is three times higher than that obtained in Ref.~\cite{Jiang:2016woz}. However, the values of the local non-relativistic vorticity on the reaction plane at $t=1$ fm$/c$ depicted in the middle panel of Fig.~\ref{fig:vort_etax} agree to a great extent in the two models, with maximum absolute values of about 2 fm$^{-1}$ in the area shown in the plot.
\\
Despite the different initial conditions and models, our results for the angular momentum, the longitudinal velocity profile (Fig.~\ref{fig:vz_x}) and the space-averaged vorticity (Fig.~\ref{fig:vort_t}) are qualitatively in agreement with previous works \cite{Deng:2016gyh, Jiang:2016woz}.

It is worth to comment on the relation between the spatial profiles of vorticity and density and the consequences on final flow observables.
Looking at the quadrupolar pattern of the local distribution of vorticity on the reaction plane shown in Fig.~\ref{fig:vort_etax} we observe that the system is rotating clockwise in the regions where $\omega_y<0$ and counterclockwise where $\omega_y>0$. As we can see from the upper panels of Fig.~\ref{fig:tilted} the regions with $\omega_y<0$ correspond to the areas with higher proper density. This is the reason why, as evident from Fig.~\ref{fig:vort_t}, the vorticity weighted over the fireball is negative and the fireball is globally rotating clockwise. As a consequence the directed flow generated in the central rapidity region has a negative slope, see Fig.~\ref{fig:v1ch_rhicb7}.

\section{Directed flow of D mesons at top RHIC energy}
\label{RHICresults}

One of the main goals of this paper is to show how the rapidity dependence of the directed flow $v_1$ of neutral $D$ mesons is originated by the vorticity of the bulk matter coming from a a tilted longitudinal distribution and the non-perturbative interaction of HQ that induce a transverse pressure push that drags the HQs in the transverse direction; on top of this  there is a charge motion induced by the electromagnetic fields produced in the collision mainly by spectator protons. We follow the ideas of Refs.~\cite{Das:2016cwd, Chatterjee:2017ahy, Chatterjee:2018lsx}.
In this section we refer to Au+Au collisions at top RHIC energy $\sqrt{\sigma_{NN}}=200$ GeV, whose experimental data for the $v_1(y)$ of $\Dz$ and $\Dzb$ mesons have been measured by the STAR Collaboration in the centrality bin 10-80\% \cite{Adam:2019wnk}.
We performed simulation at $b=9$ fm in order to compute the $v_1$ of charmed mesons and disentangle the effects of the electromagnetic fields and the vorticity.
As explained in the previous sections, angular momentum and vorticity is created in our model by means of an asymmetric initial distribution of the QGP density in the $x-\eta_s$ plane and the parameters in Eqs.~\eqref{eq:Hfunct}--\eqref{eq:f+f-} are fixed to reproduce the experimental data of the directed flow of charged particles. We use the STAR data in the centrality classes 5-40\% and 40-80\% in order to have a reference band for our simulations at $b=9$ fm.

\begin{figure}[!hbt]
\centering
\includegraphics[width=0.95\columnwidth]{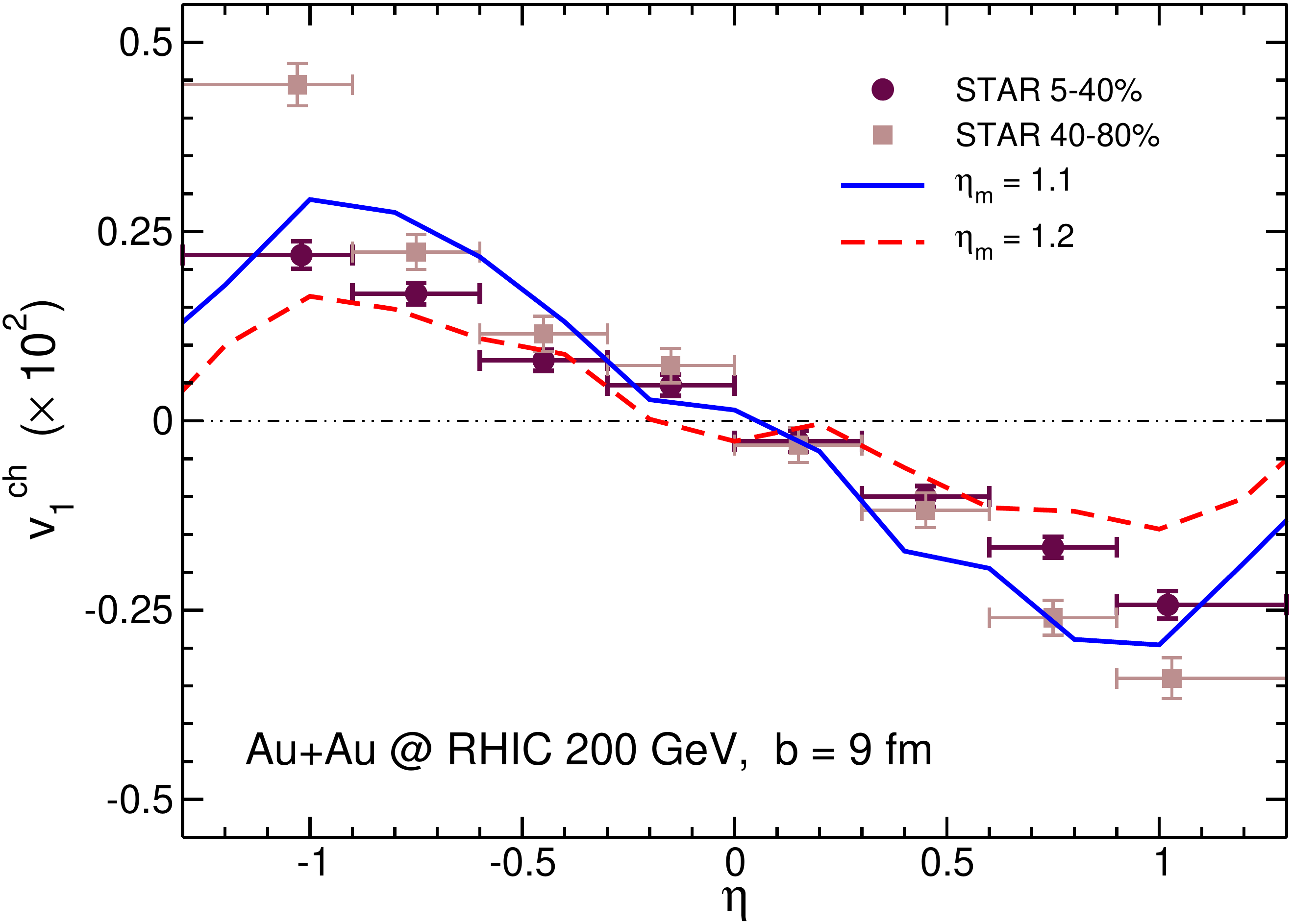}
\caption{Directed flow of charged particles versus pseudorapidity for Au+Au collisions at RHIC energy $\sqrt{\sigma_{NN}}=200$ GeV with impact parameter $b=9$ fm. The two lines corresponds to different values of the parameter $\eta_m$. The experimental data are from the STAR Collaboration \cite{Abelev:2008jga}.}
\label{fig:v1ch_rhicb9}
\end{figure}

In Fig.~\ref{fig:v1ch_rhicb9} we compare the experimental data of the directed flow of charged particles as a function of pseudorapidity with our results obtained with parameters in Eqs.~\eqref{eq:Hfunct}--\eqref{eq:f+f-} $\eta_{s0}=1$, $\sigma_{\eta}=1.3$ and two different values of $\eta_m$.
We see that the STAR points for the centrality classes 5-40\% (maroon circles) and 40-80\% (brown squares) \cite{Abelev:2008jga} are matched respectively with $\eta_m=1.2$ (dashed red line) and $\eta_m=1.1$ (solid blue line); therefore, the results of simulations with these two values of $\eta_m$ can be considered as lower and upper limits for the comparison to the experimental data of 10-80\% central collisions, that we are interested to.
From Fig.~\ref{fig:v1ch_rhicb9} we can also get information on how the initial density asymmetry introduced with Eqs.~\eqref{eq:ini_dens}--\eqref{eq:f+f-} is translated to the final azimuthal asymmetry measured by the directed flow. Indeed, a larger value of $\eta_m$ corresponds to a milder asymmetry between the positive and negative $\eta_s$ sides of the reaction plane and this leads in turn to a lower value of the particle directed flow.
It can be noticed that the dependence of the $v_1$ of the bulk is significantly dependent of the value of $\eta_m$ and allow to tune it with good accuracy. However for the aim of the paper it will be even more interesting to see that the HQ $v_1$ appears to be nearly independent on it.

\begin{figure}[!hbt]
\centering
\includegraphics[width=0.95\columnwidth]{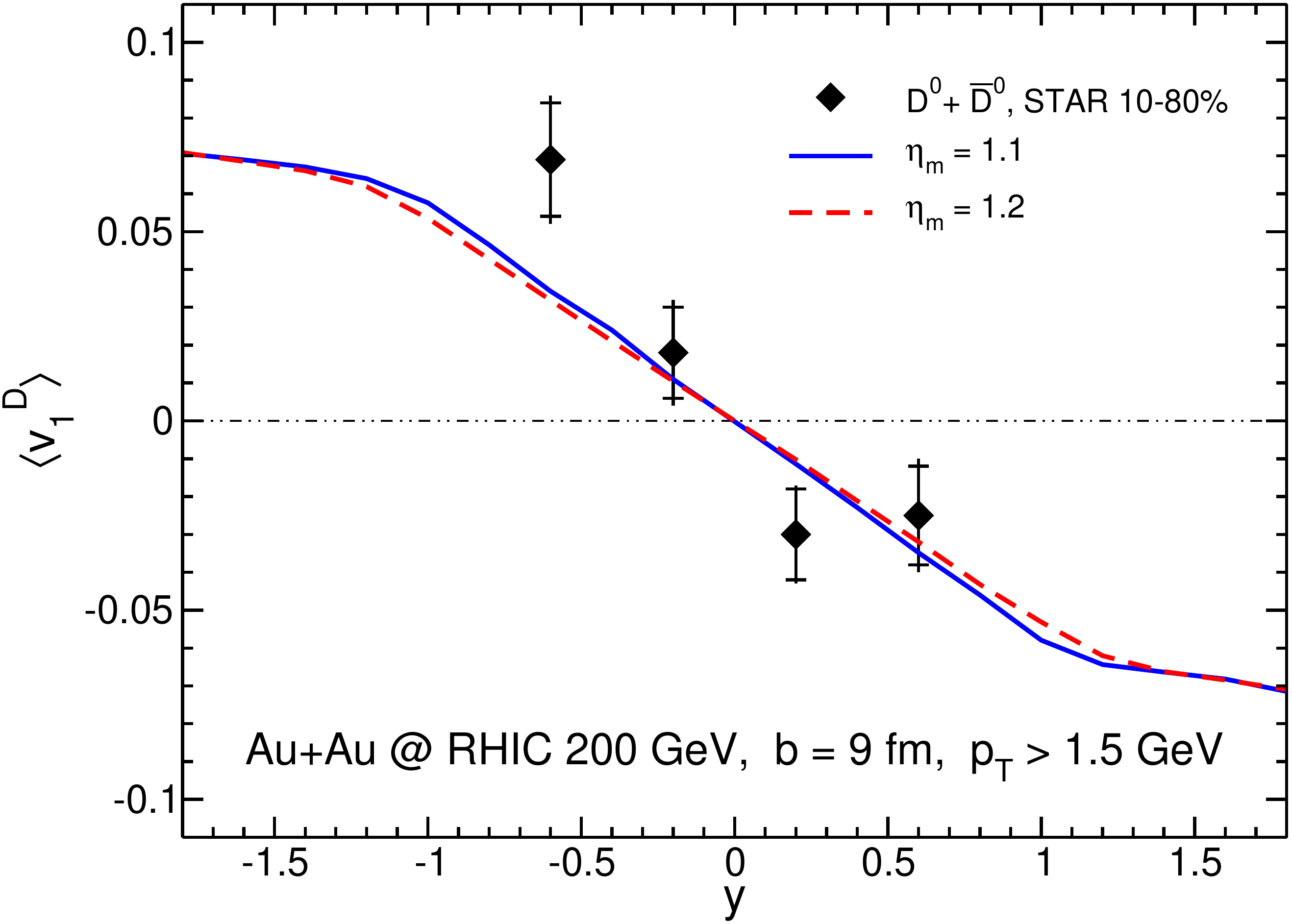}
\\[\baselineskip]
\includegraphics[width=0.95\columnwidth]{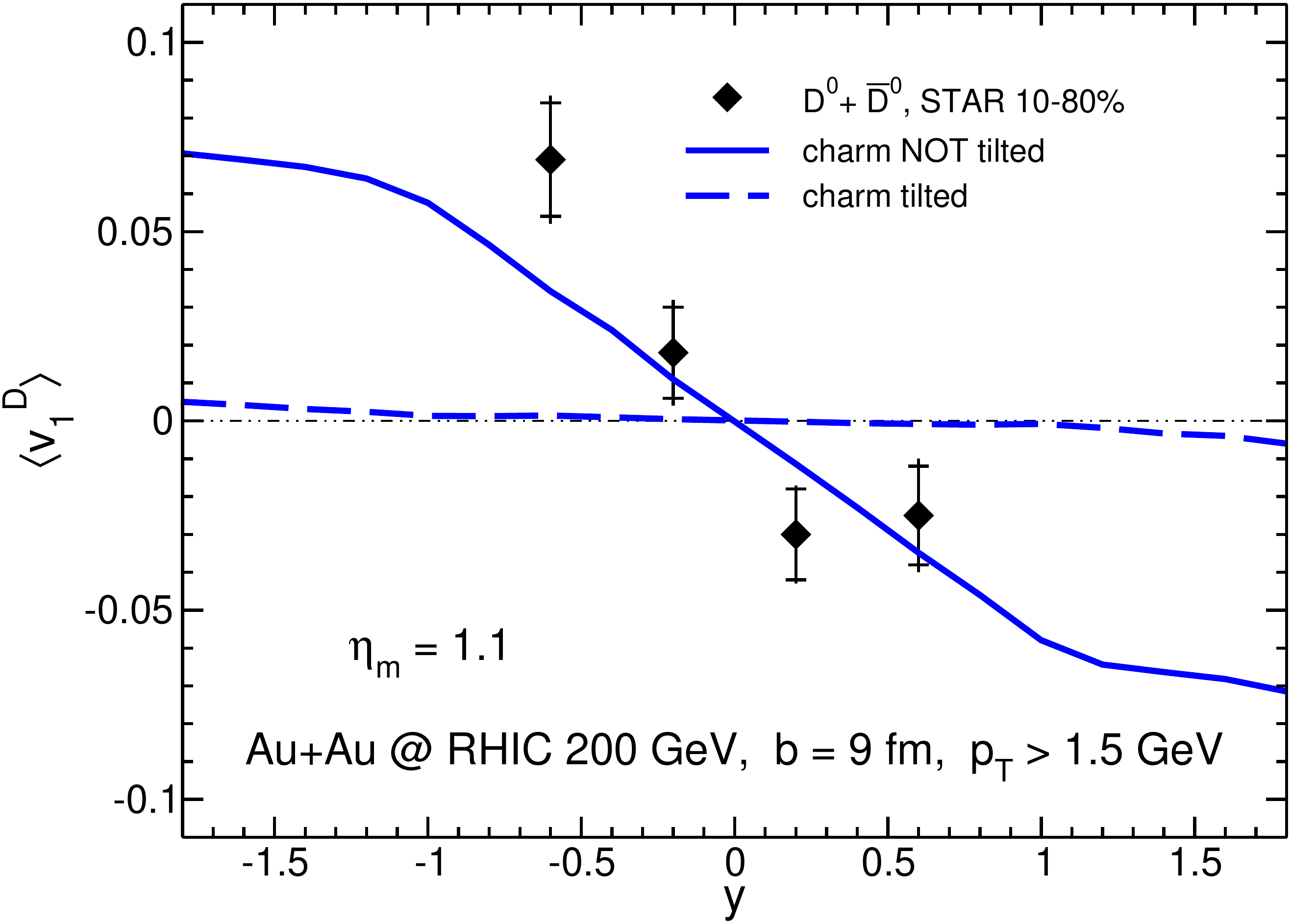}
\caption{Combined directed flow of $\Dz$ and $\Dzb$ mesons as a function of rapidity for Au+Au collisions at top RHIC energy $\sqrt{\sigma_{NN}}=200$ GeV with impact parameter $b=9$ fm in comparison with the experimental data from STAR Collaboration \cite{Adam:2019wnk}. Top panel: results for two different values of the parameter $\eta_m$: $\eta_m=1.1$ (solid blue line) and $\eta_m=1.2$ (dashed red line). Bottom panel: comparison of the result at $\eta_m=1.1$ (solid blue line) with simulations in which also charmed quarks are longitudinally distributed according to Eqs.~\eqref{eq:ini_dens}--\eqref{eq:f+f-} (dashed blue line).}
\label{fig:v1D_rhicb9_avg}
\end{figure}

In Fig.~\ref{fig:v1D_rhicb9_avg} we plot the rapidity dependence of the averaged directed flow of $\Dz$ and $\Dzb$ mesons.
In the top panel we show the result for two different values of the parameter $\eta_m$ of Eqs.~\eqref{eq:f+f-} as in Fig.~\ref{fig:v1ch_rhicb9}: the solid blue line and the dashed red line are obtained respectively with $\eta_m=1.1$ and $\eta_m=1.2$.
We notice that, even though the charged particle $v_1$ is very sensitive to the parameter $\eta_m$, the directed flow of $D$ mesons is influenced only slightly.
A far big influence on $v_1^D$ is instead exerted by the initial condition for the space distribution of charmed quarks, namely if $c$ and $\overline{c}$ are also distributed along $\eta_s$ according to Eqs.~\eqref{eq:ini_dens}--\eqref{eq:f+f-} as done for the light quarks, hence giving them since the beginning of the collision evolution an asymmetry in density and longitudinal velocity profile, see Figs.~\ref{fig:vz_x} and \ref{fig:tilted}.
The effect of the two initializations for heavy quarks is shown in the bottom panel of Fig.~\ref{fig:v1D_rhicb9_avg} for $\eta_m=1.1$ comparing simulations in which vorticity is initially given (dashed line) or not (solid line) to charm and anticharm distributions. This result is independent on $\eta_m$ (for the values explored in this works), hence an explicit representation of the lines for different $\eta_m$ is discarded.
\\
Our result for the directed flow of $D$ mesons is few times larger than that found in previous works \cite{Bratkovskaya:2004ec, Chatterjee:2017ahy} and is in the right ballpark with respect to the STAR data \cite{Adam:2019wnk} labelled with black diamonds; however, the experimental uncertainties are still quite large for a very clear comparison.
Moreover, the values of $\langle v_1^{D}\rangle$ in the central rapidity range $\vert y\vert<1$ are about 20-30 times larger than those of charged particle shown in Fig.~\ref{fig:v1ch_rhicb9}. As anticipated in Sec.~\ref{ic_vort}, this is a result of the shift in the transverse profiles of the tilted QGP fireball and the symmetrically-distributed emission points of heavy quarks from initial hard scatterings \cite{Chatterjee:2017ahy}; this shift leads to an amplification of directed flow of $D$ mesons due to the collisions undergone with the QGP and the drag exerted from the bulk matter.

\begin{figure}[!hbt]
\centering
\includegraphics[width=0.95\columnwidth]{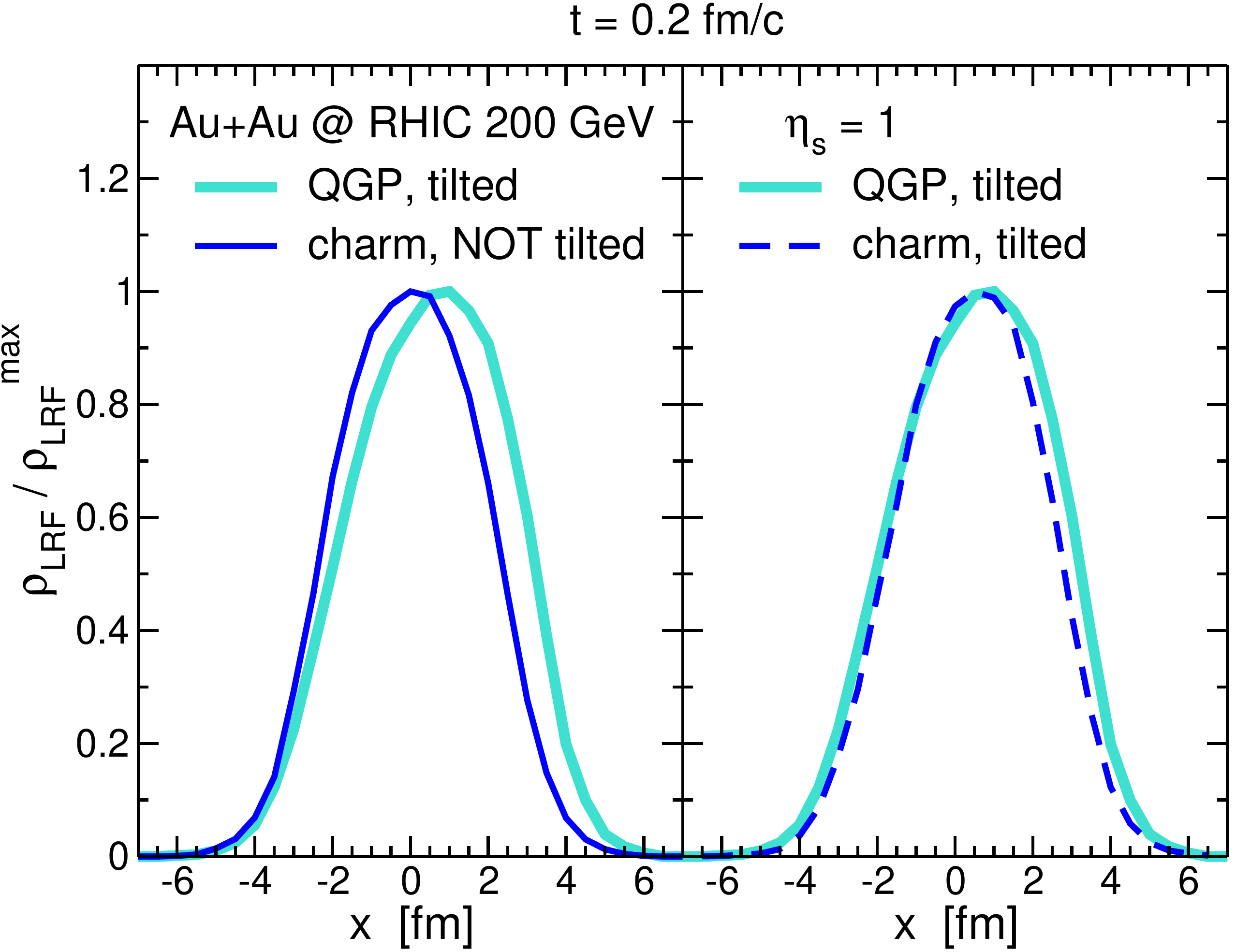}
\caption{Normalized density distributions of the bulk matter (thick turquoise lines) and charm quarks (thin blue lines) at $\eta_s=1$ at the initial time of the simulation; the solid blue line corresponds to initial distribution of charms not tilted and the dashed blue line to initial tilted charm distribution.}
\label{fig:rhox_QGP_HQ}
\end{figure}

However, if the $c-\overline{c}$ pairs are initialized according with a tilted source, as done for gluons and light quarks, the transverse spatial distributions of heavy quarks and quark-gluon plasma are superimposed. We show this in Fig.~\ref{fig:rhox_QGP_HQ} where the density profile
of both the bulk and the charm quarks are shown as a function of the transverse coordinate $x$ at the space rapidity $\eta_s=1$.
Once the charm are distributed as the bulk matter the large $v_1$ of the HQs disappears. In fact in this case
the small $v_1$ comes only by the weak rotation as for the bulk matter and hence this leads to a $v_1$ even smaller then the one of the bulk, as shown by the dashed line in the bottom plot of Fig.~\ref{fig:v1D_rhicb9_avg}. 
This allows us to understand a fundamental aspect of the directed flow. Its origin appears to be different from the one of light particles. The large $v_1$ is due to a pressure push of the bulk that having a different space distribution pushes the heavy quarks toward the negative $x-$direction at positive $\eta_s$.

\begin{figure}[!hbt]
\centering
\includegraphics[width=0.95\columnwidth]{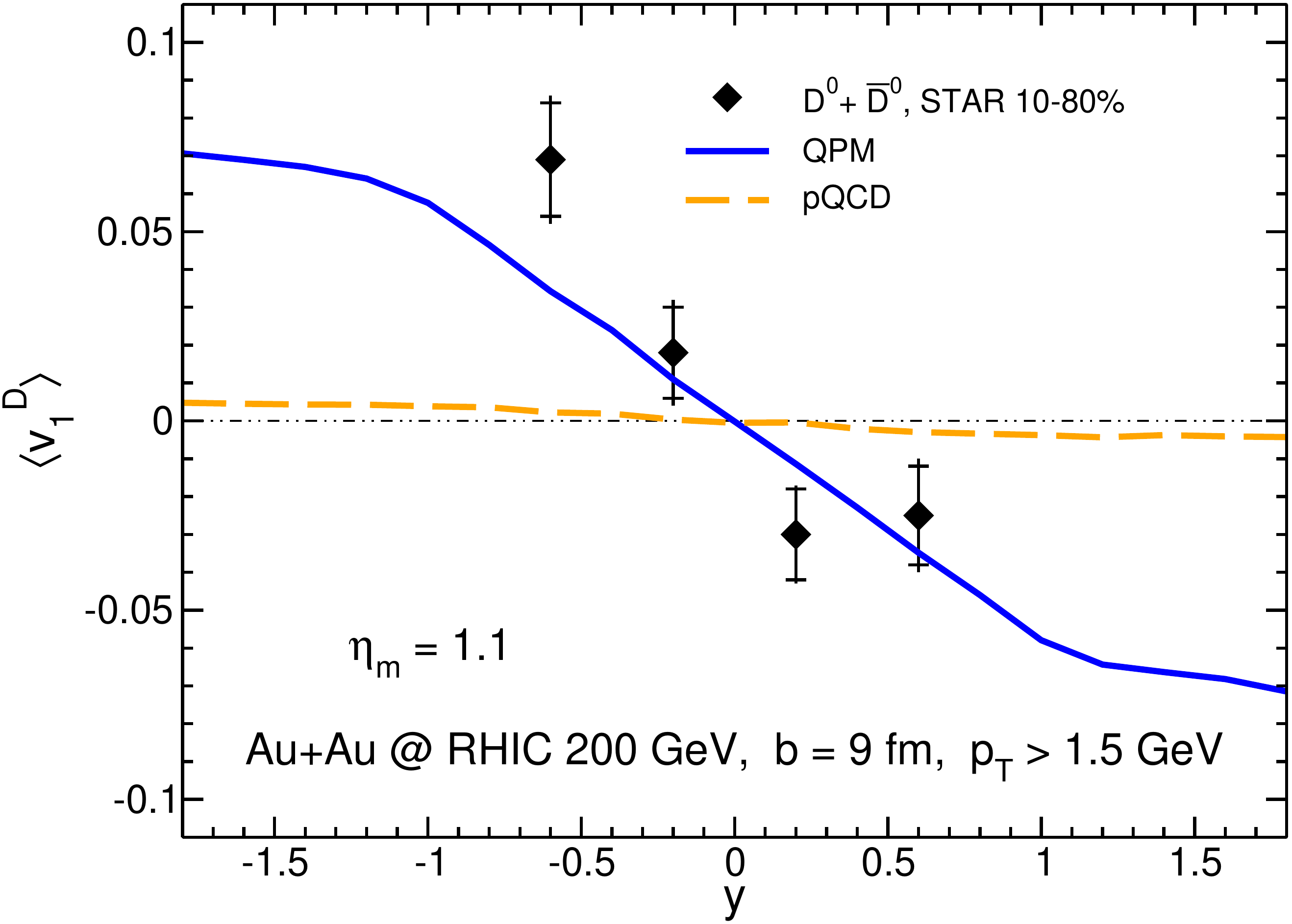}
\caption{Combined directed flow of $\Dz$ and $\Dzb$ mesons as a function of rapidity for Au+Au collisions at top RHIC energy $\sqrt{\sigma_{NN}}=200$ GeV with impact parameter $b=9$ fm. The blue solid line corresponds to the same calculation as in Fig.~\ref{fig:v1D_rhicb9_avg} for charm that are non-tilted and with a drag according to the non-perturbative QPM model, the dashed orange line is the result employing a drag according to pQCD.}
\label{fig:v1D_QPM_pQCD}
\end{figure}
The effect of such a pressure gradient in the transverse direction is however effective only because the interaction
of the HQ with the bulk is largely non-perturbative. To show this we plot in Fig.~\ref{fig:v1D_QPM_pQCD} the $v_1(y)$ 
if one keeps the non-tilted space distribution but assume a drag and diffusion of the charm quarks according to pQCD. It is clear that in this case the bulk is not able to transmit the transverse push because the drag is much smaller. Therefore in turn the large $v_1$ of $D$ meson is a signature of the non-perturbative interaction of the HQ with the bulk hot QCD matter that is however visible only because the bulk matter is tilted in the longitudinal direction.
We get a reasonably good prediction of $v_1^D$ because we assume for heavy quarks the drag and diffusion extrapolated from the phenomenological analysis of $R_{AA}(p_T)$ and $v_2(p_T)$ as summarized in Ref.~\cite{Dong:2019unq}.

\begin{figure}[!hbt]
\centering
\includegraphics[width=0.95\columnwidth]{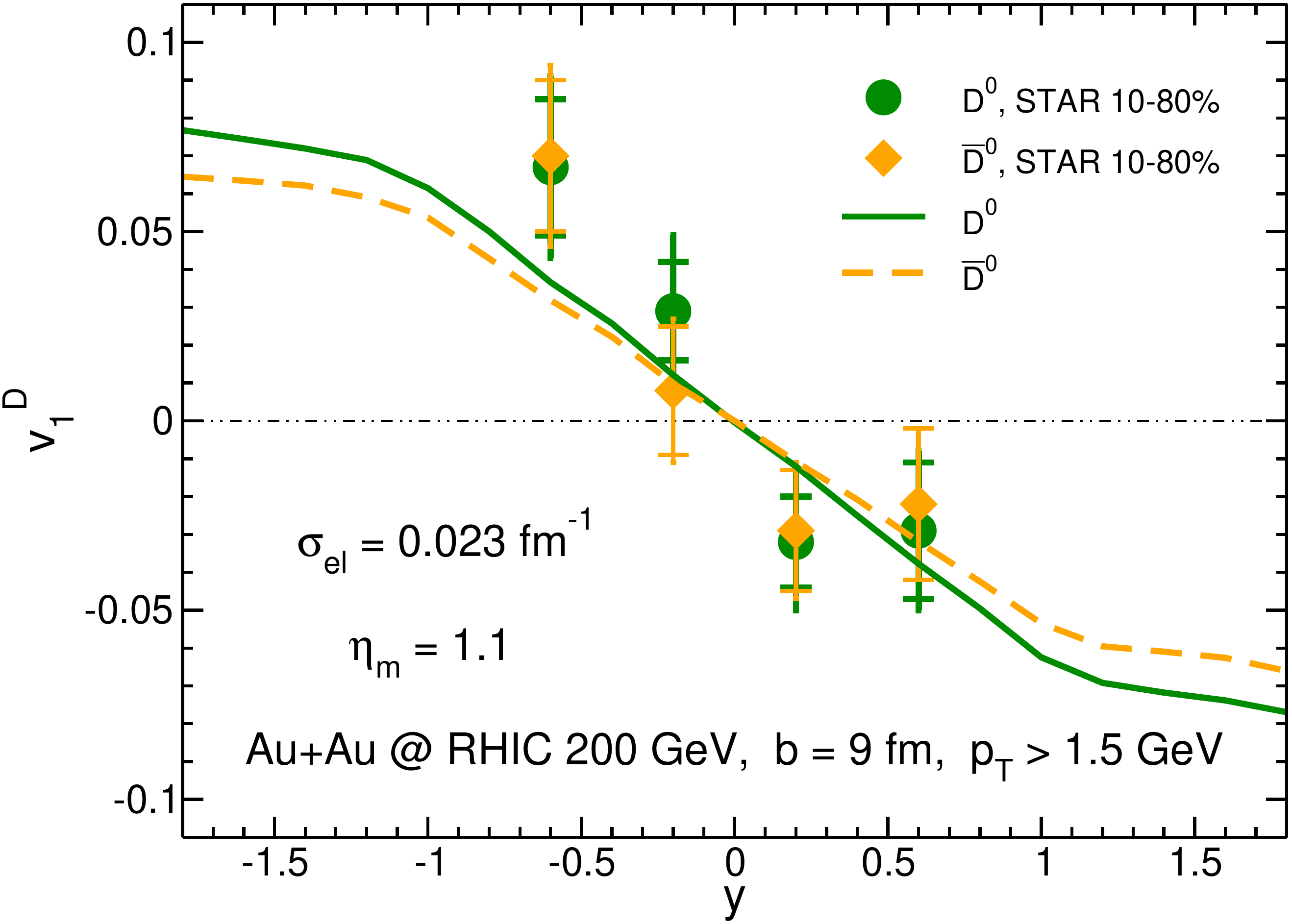}
\caption{Directed flow of $\Dz$ and $\Dzb$ mesons as a function of rapidity for Au+Au collisions at RHIC energy $\sqrt{\sigma_{NN}}=200$ GeV with impact parameter $b=9$ fm. The experimental data from STAR Collaboration \cite{Adam:2019wnk} are also shown.}
\label{fig:v1D_rhicb9_nyEMF}
\end{figure}

In Ref.~\cite{Das:2016cwd} some of us pointed out for the first time that the electromagnetic field generated in the early stage of the heavy-ion collisions can induce a splitting in the $v_1$ between charm and anti-charm. The calculation were performed in a Langevin framework discarding the longitudinal tilted distribution of the bulk. We present here the first results 
incorporating the electromagnetic field dynamics in our Boltzmann transport approach, as described in Sec.~\ref{transport}, including the vorticity and the tilted bulk distribution.
In Fig.~\ref{fig:v1D_rhicb9_nyEMF} it is shown the influence of the electromagnetic fields on the rapidity dependence of the directed flow of $\Dz$ and $\Dzb$ mesons for Au+Au collisions at top RHIC energy with impact parameter $b=9$ fm. The simulations include the Lorentz force in the Boltzmann equation and therefore the effect of the electromagnetic fields on particle dynamics.
For the computation of the time evolution of the electric and magnetic fields the value of the electric conductivity used is $\sigma_{el}=0.023$ fm$^{-1}$, a central value from lattice QCD calculation at $T=2\,T_c$.
We have performed also simulations without the electromagnetic fields and found that the two lines of the $v_1$ of $\Dz$ and $\Dzb$ lie one on top of each other.
We see clearly from Fig.~\ref{fig:v1D_rhicb9_nyEMF} that a splitting of the two curves is generated when the electromagnetic fields are considered \cite{Das:2016cwd, Chatterjee:2018lsx}: if we focus on the region $y>0$ the $v_1$ of $\Dz$ is pushed downwards and that of $\Dzb$ is pushed upwards.
Since we are assuming that $\Dz$ and $\Dzb$ are produced by fragmentation of $c$ and $\overline{c}$ quarks respectively, the $v_1$ of neutral $D$ mesons is  driven by the $v_1$ of charm quarks and antiquarks; from Fig.~\ref{fig:v1D_rhicb9_nyEMF} we deduce that at forward rapidity positively-charged particles ($c$) get a negative contribution from the electromagnetic fields to the $v_1$ and negatively-charged particles ($\overline{c}$) get a positive contribution -- and vice versa at backward rapidity. This means that the total effect coming from $B_y$ and $E_x$ which push particles in opposite directions is slightly dominated by the electric field.
This is also consistent with the results obtained in previous studies for light mesons, both in hydrodynamic calculations of RHIC Au+Au and LHC Pb+Pb collisions \cite{Gursoy:2014aka, Gursoy:2018yai} and in transport simulations of asymmetric collisions \cite{Voronyuk:2014rna, Toneev:2016bri} and small systems \cite{Oliva:2019kin} at top RHIC energy.
\\
In addition to the $v_1$ splitting induced by the electromagnetic field, there may be a further mismatch of the directed flow of $D$ mesons conveyed by the initial orbital angular momentum of the two colliding nuclei, which contributes to the directed flow of $u$ and $d$ quarks \cite{Dunlop:2011cf, Adamczyk:2017nxg, Oliva:2019kin}: as a consequence, the $\Dzb$ ($\overline{c}u$ system) gets a larger $v_1$ at forward rapidity and a smaller $v_1$ at backward rapidity, following the directed flow pattern of spectators. This effect, that is not included in our model, would increase the splitting between $\Dz$ and $\Dzb$ especially at higher $y$, while in the central rapidity region the force exerted by the electromagnetic field dominates \cite{Oliva:2019kin}. However it can be expected to give correction of the order of $10^{-3}$
on $v_1$ given the small transverse flow of the bulk.

\begin{figure}[!hbt]
\centering
\includegraphics[width=0.95\columnwidth]{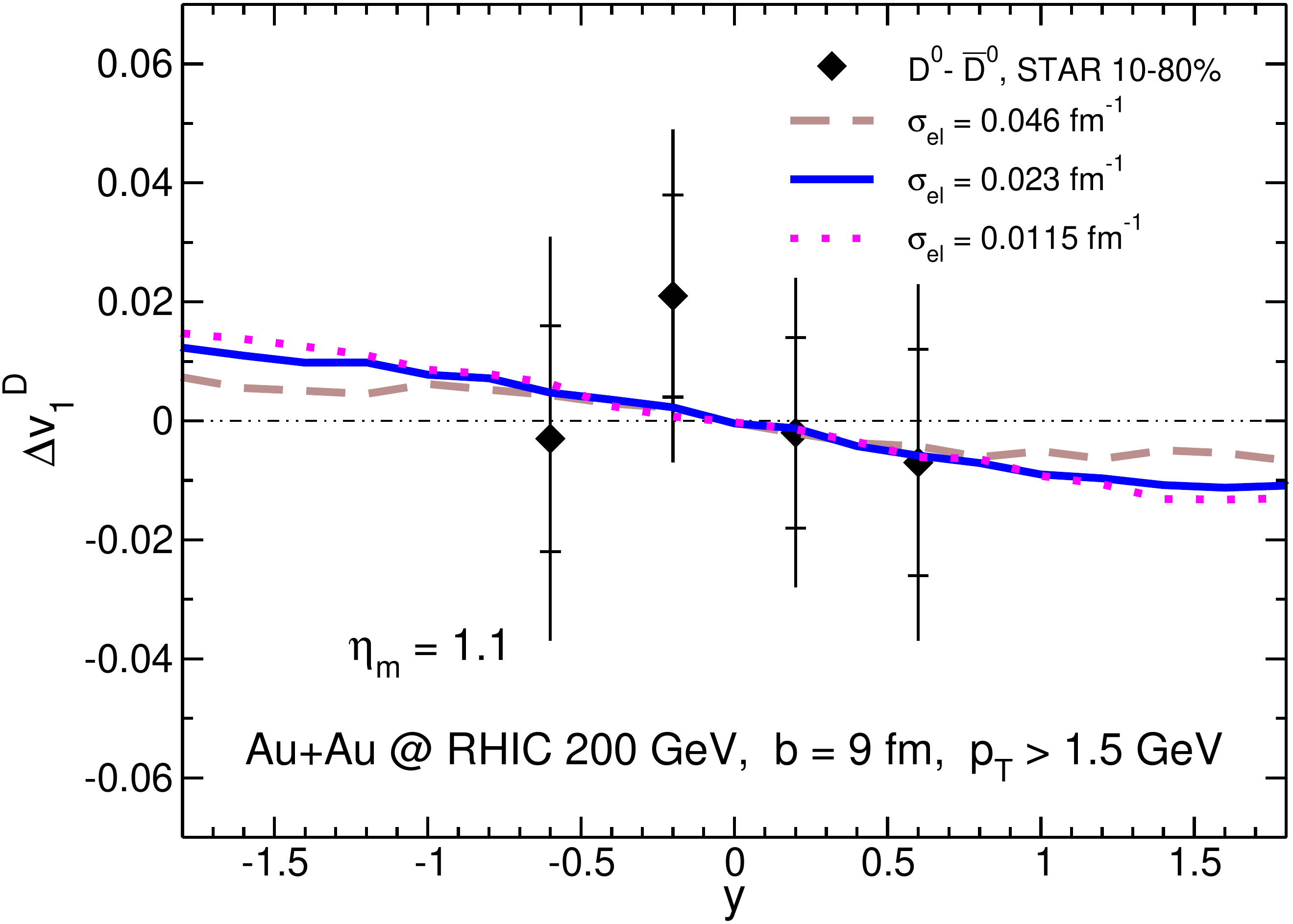}
\caption{Rapidity dependence of the directed flow difference $\Delta v_1$ between $\Dz$ and $\Dzb$ mesons at top RHIC energy $\sqrt{\sigma_{NN}}=200$ GeV for different values of the electric conductivity $\sigma_{el}$. The experimental data are from the STAR Collaboration \cite{Adam:2019wnk}.}
\label{fig:v1D_rhicb9_diff}
\end{figure}

We have also investigated how the electric conductivity affect the evolution of the electromagnetic fields and, consequently, the $v_1$ splitting of $D$ mesons. The last one can be easily quantified by the difference in the directed flow between one particle and the corresponding antiparticle, i.e., for $D$ mesons $\Delta v_1^D=v_1(\Dz)-v_1(\Dzb)$. 
This quantity is plotted in Fig.~\ref{fig:v1D_rhicb9_diff} for different values of the electric conductivity: the result for $\sigma_{el}=0.023$ fm$^{-1}$ (solid blue line), which is the same of the solid lines in Fig.~\ref{fig:v1D_rhicb9_nyEMF}, is compared with those obtained with $\sigma_{el}=0.0115$ fm$^{-1}$ (dotted magenta line) and $\sigma_{el}=0.46$ fm$^{-1}$ (dashed brown line), which correspond respectively to halving and doubling the value of electric conductivity firstly considered. Only in the latter case, larger $\sigma_{el}$, 
we see a mild change in the $\Delta v_1^D$ for higher rapidities $\vert y\vert>1$.
This can be understood by looking at the middle and right panels of Fig.~\ref{fig:EMF_rhicb9} in which we show the temporal behaviour at forward space-time rapidity of the fields for the considered values of electrical conductivity: for $\sigma_{el}=0.46$ fm$^{-1}$ the magnetic field has a milder decrease with respect to smaller conductivities and this in turn means that the produced electric field is lower and then its effect on the propagation of particle is smaller; this explains why the splitting $\Delta v_1^D$ is smaller with respect to that obtained with lower $\sigma_{el}$. 

\begin{figure}[!hbt]
\centering
\includegraphics[width=0.95\columnwidth]{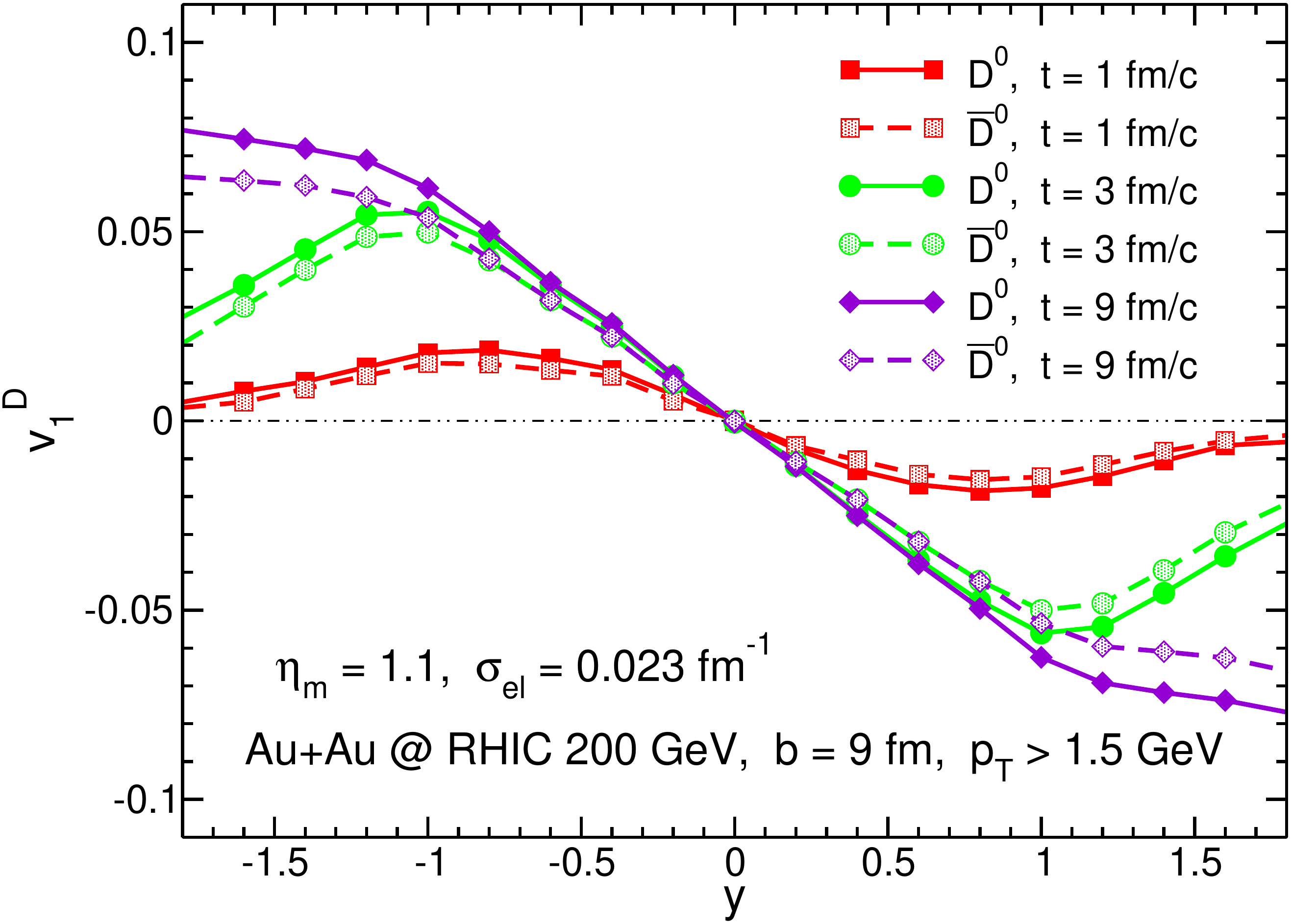}
\caption{Time evolution of the directed flow of $\Dz$ and $\Dzb$ as a function of rapidity at for Au+Au collisions at top RHIC energy $\sqrt{\sigma_{NN}}=200$ GeV with impact parameter $b=9$ fm.}
\label{fig:v1D_rhicb9_time}
\end{figure}

\begin{figure}[!hbt]
\centering
\includegraphics[width=0.95\columnwidth]{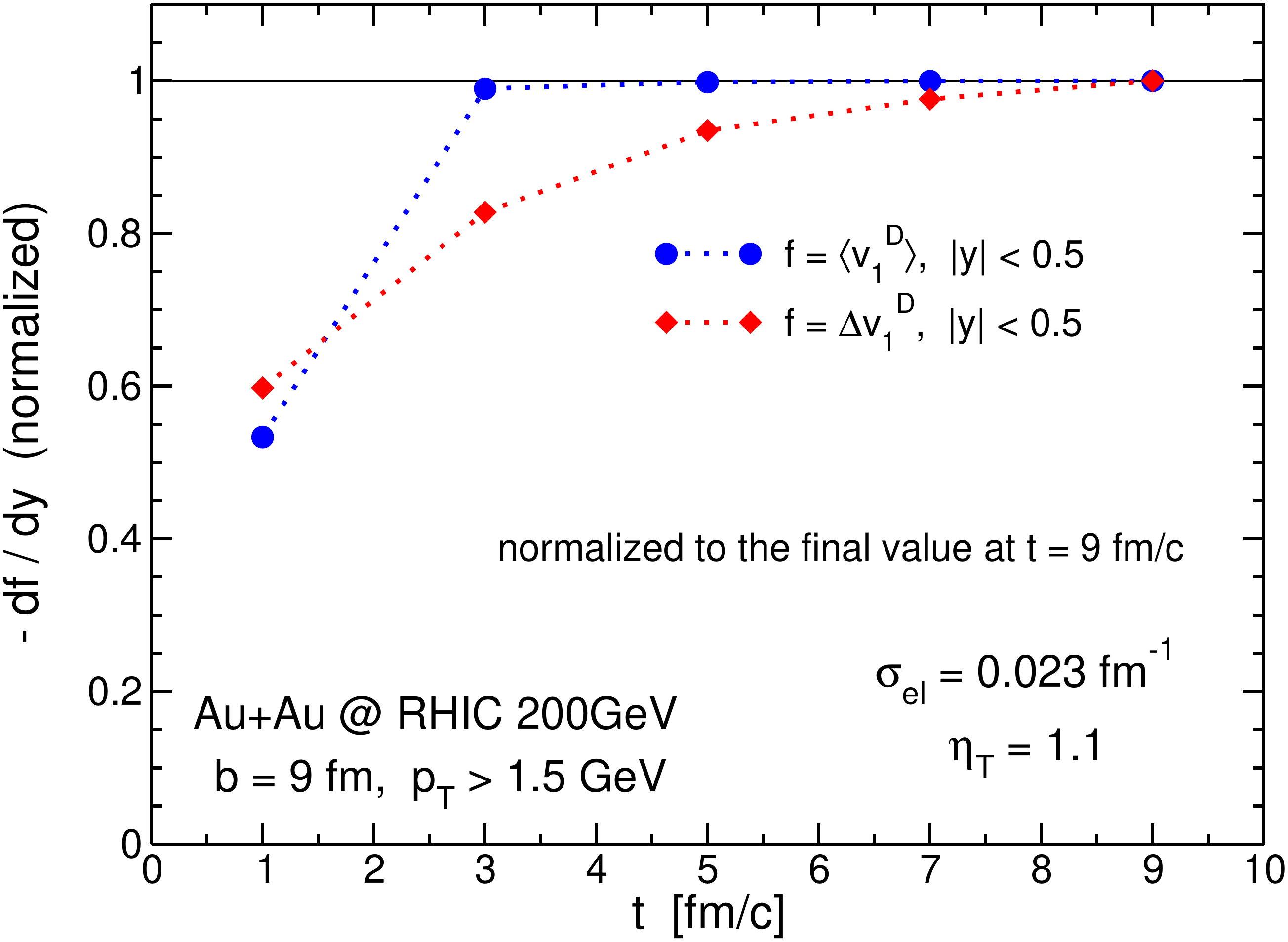}
\caption{Time evolution of the slope of the averaged directed flow and the splitting of neutral $D$ mesons as a function of rapidity at RHIC energy $\sqrt{\sigma_{NN}}=200$ GeV. The values are normalized to their maximum for each quantity.}
\label{fig:v1_dv1_time_evol}
\end{figure}

In Fig.~\ref{fig:v1D_rhicb9_time} we plot the time evolution of the $v_1$ of $\Dz$ and $\Dzb$ versus rapidity for simulations with $\sigma_{el}=0.023$ fm$^{-1}$. We notice that at midrapidity most of the $v_1$ of $\Dz$ and $\Dzb$ is generated within the first 3 fm$/c$; this holds for both the average and the difference of the $D$ meson $v_1$, which give information on the temporal behaviour of the imprint of the vorticity and the electromagnetic field respectively.
In particular, in Fig. \ref{fig:v1_dv1_time_evol} we show that for the $y-$dependence of both $\langle v_1^D\rangle$ and $\Delta v_1^D$ the slope in the central rapidity region $\vert y\vert<0.5$ reaches half of its maximum (absolute) value in the first 1 fm$/c$; then, the averaged $D$ meson directed flow completely saturates at 3 fm$/c$ while for the splitting there is still a mild effect of about a $20\%$ at later times.

\section{Directed flow of D mesons at top LHC energy}
\label{LHCresults}

\begin{figure}[!hbt]
\centering
\includegraphics[width=0.95\columnwidth]{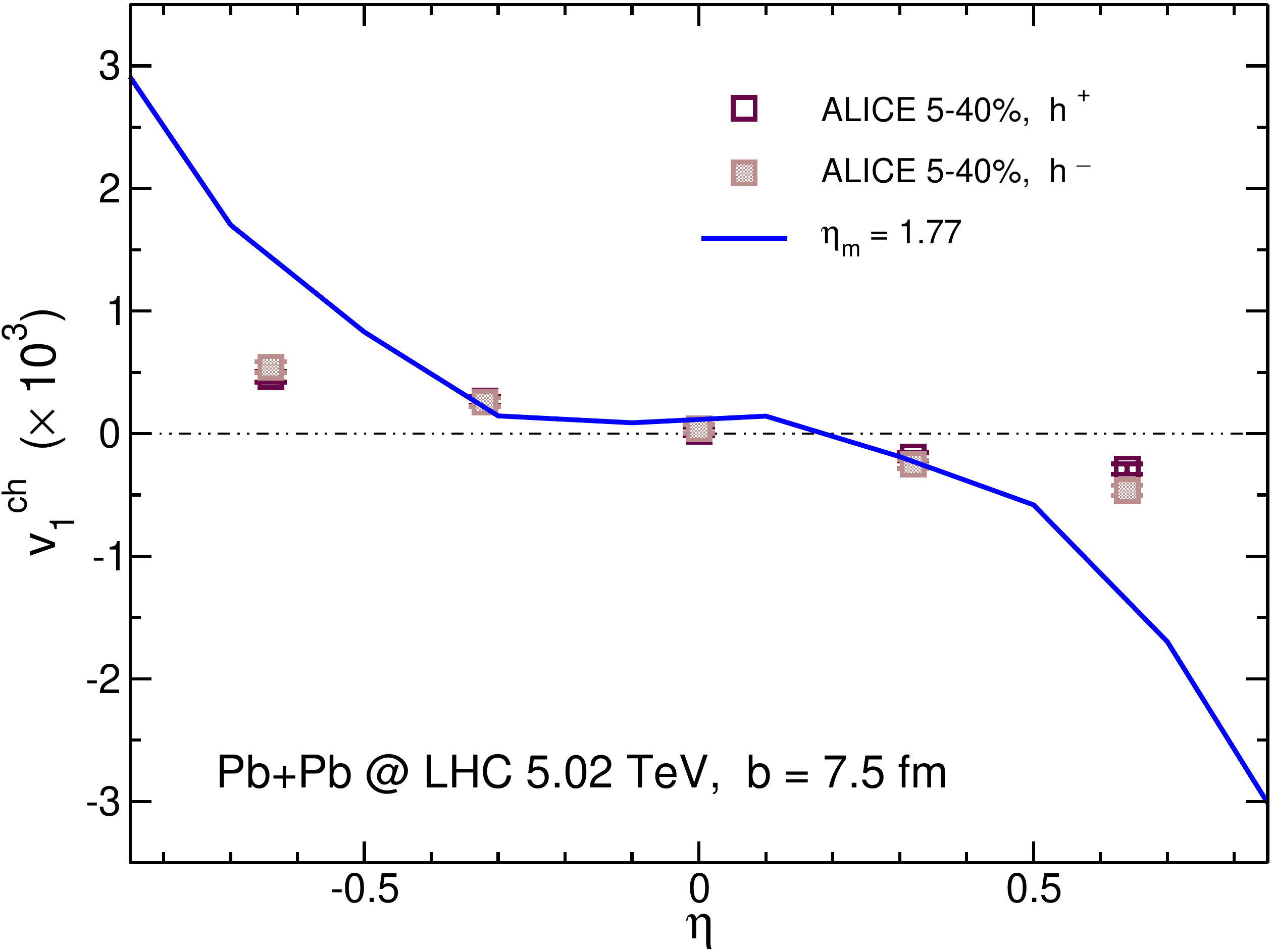}
\caption{Directed flow of charged particles as a function of pseudorapidity for Pb+Pb collisions at LHC energy $\sqrt{\sigma_{NN}}=5.02$ TeV with impact parameter $b=7.5$ fm; the experimental data are from the ALICE Collaboration \cite{Acharya:2019ijj}.}
\label{fig:v1ch_lhc}
\end{figure}

In this section we study the directed flow of $D$ mesons in Pb+Pb collisions at LHC energy $\sqrt{\sigma_{NN}}=5.02$ TeV; the ALICE Collaboration has recently published the experimental measurements of the $v_1(\eta)$ of positively and negatively charged particles as well as neutral $D$ mesons respectively in the centrality classes 5-40\% and 10-40\% \cite{Acharya:2019ijj} which correspond to $b=7.5$ fm.
We performed simulations at this value of the impact parameter both with the standard initial conditions and with the initialization that generate the vorticity in the QGP fireball.
The parameters in Eqs.~\eqref{eq:Hfunct}--\eqref{eq:f+f-} are fixed in order to match the experimental data of the $v_1$ of charged particles, as shown in Fig.~\ref{fig:v1ch_lhc}; we found $\eta_{s0}=2.0$, $\sigma_\eta=1.3$ and $\eta_m=1.77$.

\begin{figure}[!hbt]
\centering
\includegraphics[width=0.95\columnwidth]{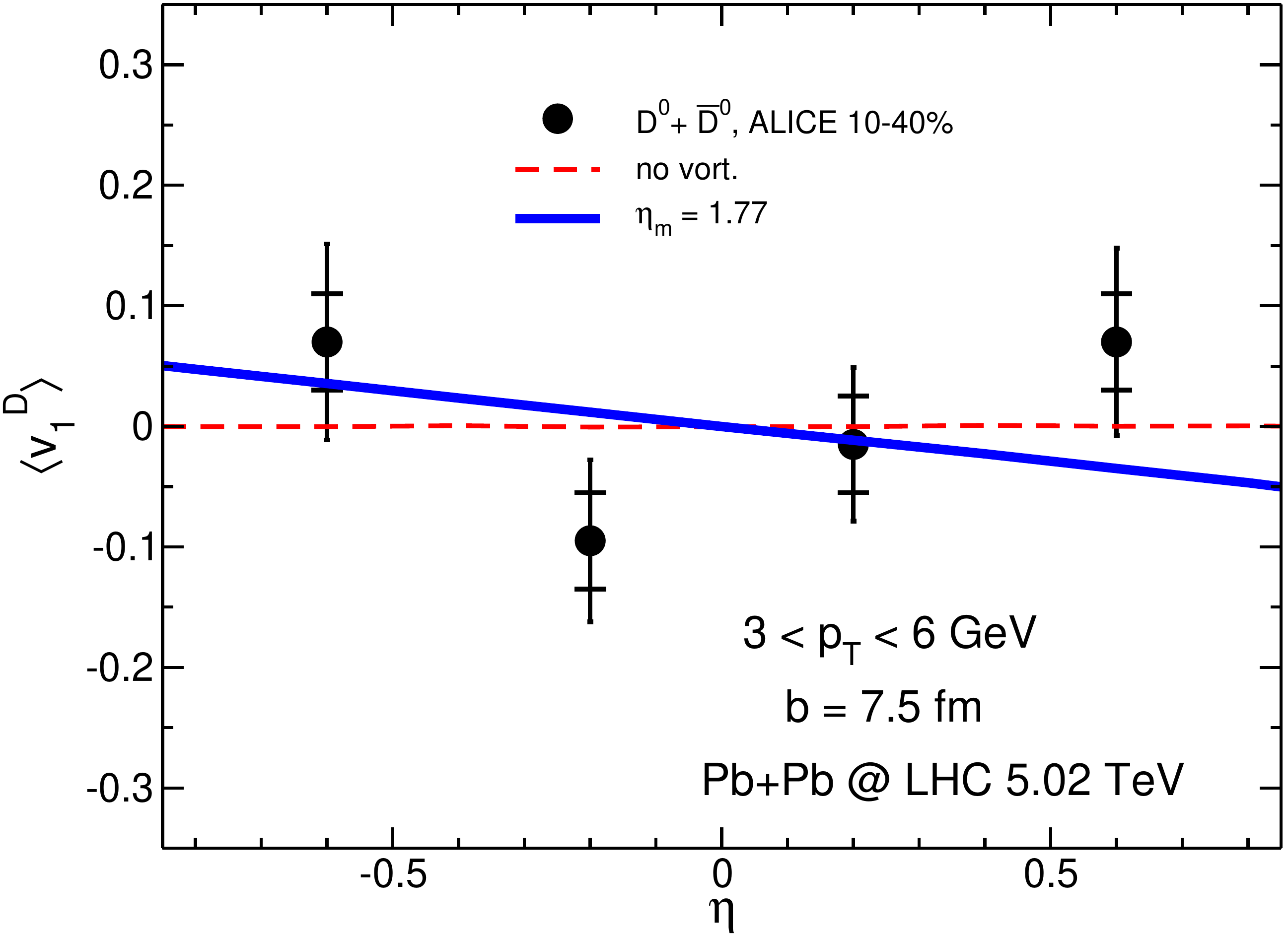}
\\[\baselineskip]
\includegraphics[width=0.95\columnwidth]{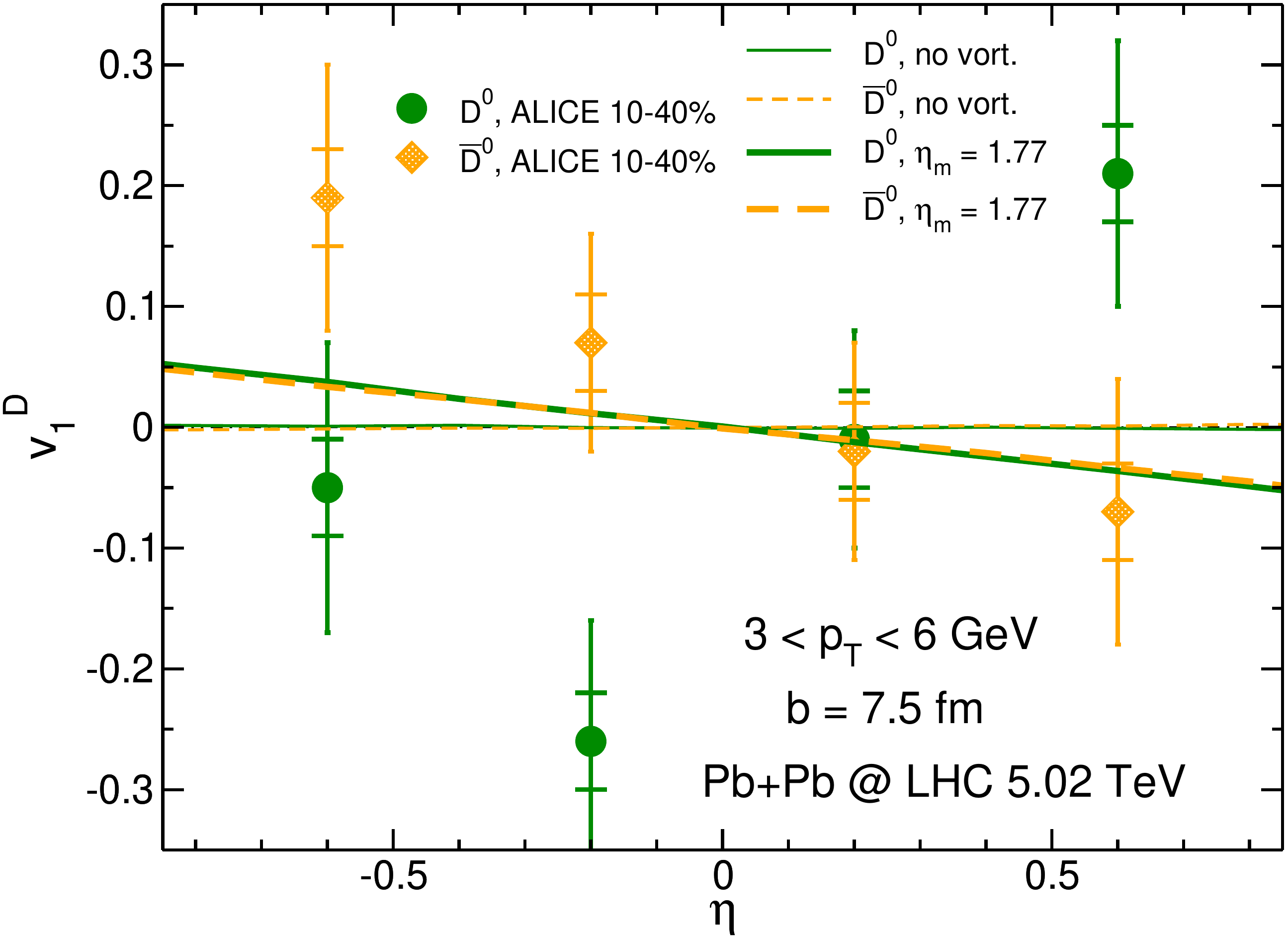}
\caption{Directed flow of $\Dz$ and $\Dzb$ mesons as a function of pseudorapidity for Pb+Pb collisions at LHC energy $\sqrt{\sigma_{NN}}=5.02$ TeV with impact parameter $b=7.5$ fm in comparison with the experimental data from the ALICE Collaboration \cite{Acharya:2019ijj}. Upper panel: experimental data and theoretical predictions for $\Dz+ \Dzb$. Lower panel: as in the upper panel for $\Dz$ and $\Dzb$ separately.}
\label{fig:v1D_lhc}
\end{figure}

The corresponding results for the $\Dz$ and $\Dzb$ are plotted in Fig.~\ref{fig:v1D_lhc} in comparison with the experimental data from ALICE Collaboration \cite{Acharya:2019ijj}. Thin lines correspond to the simulation with the standard initialization (which generate a zero directed flow of charged particles) whereas thick curves are the results with initial conditions given by Eqs.~\eqref{eq:ini_dens}--\eqref{eq:f+f-}.
Both simulations includes the electromagnetic fields,
which generate a splitting in the $v_1(\eta)$ of $D$ mesons: at forward rapidity $\Dz$ (solid green line) has a lower directed flow than its antiparticle (dashed orange line) and the opposite trend is seen at backward rapidity. This pattern is the same of that found for RHIC collisions, indicating again that the contribution of the electric field $E_x$ in the Lorentz force 
dominate over that produced by the magnetic field $B_y$.\\
Though with big uncertainties, the ALICE data indicate a positive slope for the combined directed flow of $\Dz$ and 
$\Dzb$ that is smaller then the one observed at RHIC and in agreement with the predictions of our approach.
However it is certainly required to have experimental data with much higher precision to draw any conclusions.

\begin{figure}[!hbt]
\includegraphics[width=0.95\columnwidth]{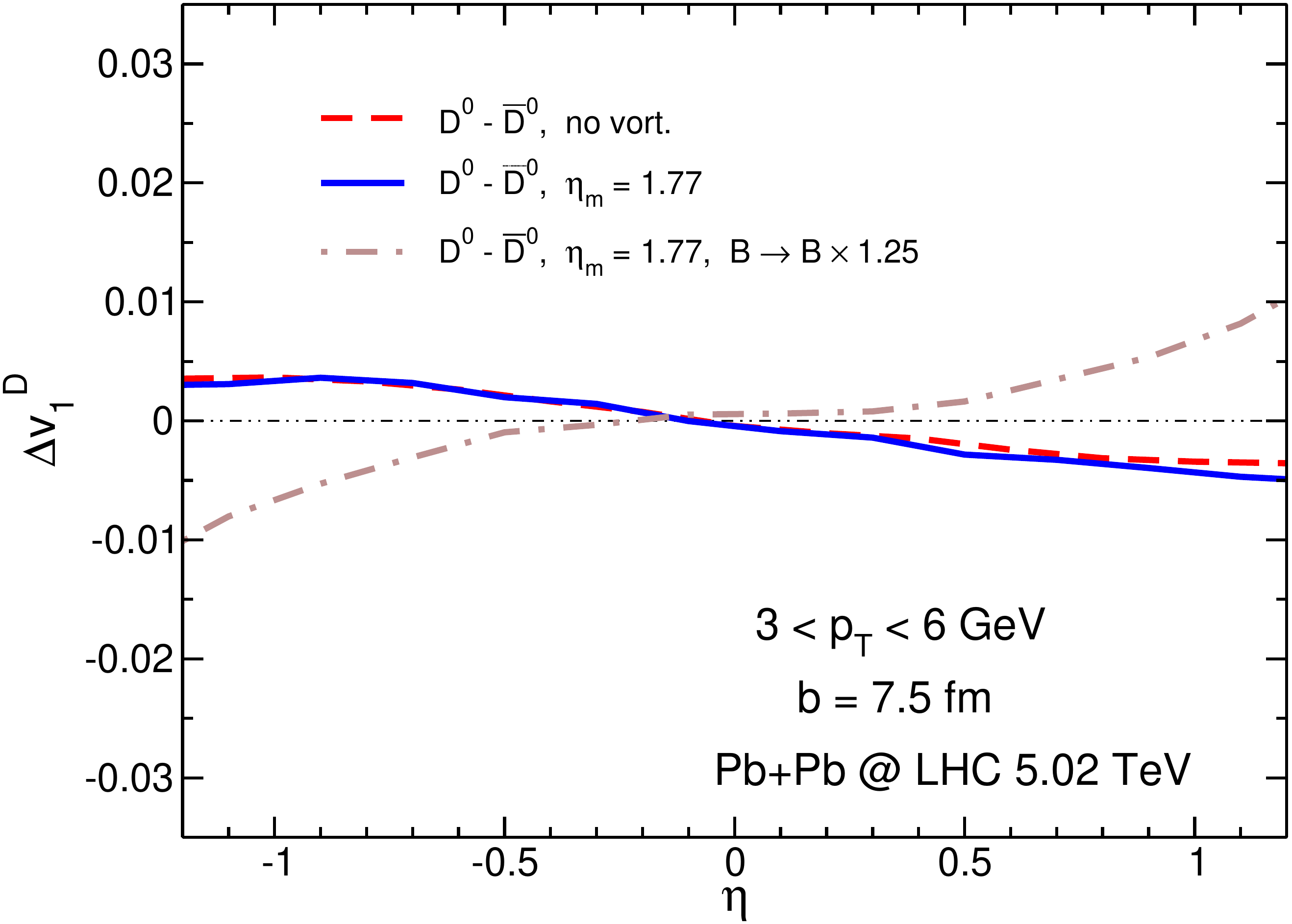}
\caption{Splitting in the directed flow of $\Dz$ and $\Dzb$ mesons as a function of pseudorapidity for Pb+Pb collisions at LHC energy $\sqrt{\sigma_{NN}}=5.02$ TeV with impact parameter $b=7.5$ fm.}
\label{fig:v1D_lhc_diff}
\end{figure}

Besides the slope of the $v_1(\eta)$ of the two particles that can be connected also to the vorticity of the fireball, the measurements of ALICE Collaboration for the splitting $\Delta v_1^D=v_1(\Dz)-v_1(\Dzb)$ show an opposite trend with respect to the STAR data for Au+Au collisions at $\sqrt{\sigma_{NN}}=200$ GeV and to our calculations at both RHIC and LHC energies.

Our result for the splitting in $\sqrt{\sigma_{NN}}=5.02$ TeV Pb+Pb collisions is shown in Fig.~\ref{fig:v1D_lhc_diff} for simulations with impact parameter $b=7.5$ fm: we found that $\Delta v_1^D$ is negative at forward rapidity and positive at backward rapidity while the ALICE experimental data have the opposite trend.
The same disagreement in sign has been found by other theoretical calculations for the charge-odd directed flow of light \cite{Gursoy:2018yai} and heavy mesons \cite{Chatterjee:2018lsx}.
Not only the sign of the splitting, but also its magnitude in absolute value measured by ALICE cannot be explained by our model: we found a $\vert\Delta v_1^D\vert$ that is two order of magnitude smaller than the experimental data.
Moreover, from Fig.~\ref{fig:v1D_lhc_diff} we see a comparable splitting in both simulations with and without vorticity, indicating that the $\Delta v_1^D$ is not affected by the initialization used for describing the effect of the angular momentum of the system.

The understanding of the ALICE measurements of the charge-dependent directed flow of $D$ mesons represents a challenge from the theoretical point of view, being them is contrast with model calculations but also with theoretical and experimental results at lower center-of-mass energy.
Possible explanations may deal with the very early time dynamics of the collision, that is the less understood stage especially at very high energy.
Even the evolution of the electromagnetic themselves is not clear in the first fractions of fm$/c$ when the system could be constituted mainly of color fields or strings and the QGP matter that sustain the electromagnetic field evolution with its electrical conductivity may be not present yet. But charm quark and antiquarks, produced in initial hard scatterings, would be witnesses of the early dynamics and of the vortical and electromagnetic fields pertinent to it.

In order to see that even a mild modification of the latter could give an insight into how the splitting of $\Dz$ and $\Dzb$ directed flow could change sign, we modified by hand the $B_y$ component of the electromagnetic field, rising by 25\% its value in each space-time cell, and keeping unchanged the values of the electric field.
The outcome of this trial is represented by the dot-dashed brown line in Fig.~\ref{fig:v1D_lhc_diff} for the splitting of the $D$ meson directed flow at LHC energy.
The effect of the increasing the magnetic field is not only to flip the sign of $\Delta v_1^D$, going in the right direction to reproduce the trend indicated by the ALICE data, but also to generated a $|\Delta v_1^D|$ even larger.
This means that a much detailed investigation of the space-time profile of the electromagnetic field has to be performed.
Recently in Ref.~\cite{Sun:2020wkg} in a schematic calculation within the Langevin approach that neglected
 the vortical dynamics and the tilted longitudinal distribution it has been shown that a magnetic field evolving slowly with respect to
 to modellings of a medium in equilibrium at constant electric conductivity can lead to a sign change of $\Delta v_1^D$
 and strength comparable to the one observed at ALICE.

\section{Conclusions}
We have presented a study of the build-up of the directed flow of charm quarks at both RHIC and LHC energy.
The predictions have been obtained by means of realistic simulation of the uRHICs based on relativistic Boltzmann transport equation
evolving on initial conditions that take into account both the vorticity of the created matter and its tilted longitudinal distribution
as well as the action of the electromagnetic fields.

The origin of the directed flow for the $D$ meson appears to be different with respect to the one of the bulk matter.
It is, in fact, not coming from the quite weak effect of rotation induced by the initial angular momentum
of the spectators transferred to the created bulk matter. 
The surprising large $v_1(y)$ observed for $\Dz$ and $\Dzb$ mesons, about 30 times larger than the one of the light charged mesons, originates from the pressure push of the bulk matter in the transverse direction. This is possible only if the peak of the bulk matter is tilted with respect to the $x-$axis and to the charm distribution.We have pointed out that such a mechanism is effective only if the interaction of HQ in the hot QCD matter is non-perturbative, otherwise the bulk matter would not be able to transfer such a push in the transverse direction to the charm quarks. 
In particular, we have show that assuming perturbative QCD drag and diffusion for the charm would lead to a $v_1(y)$ at least one order of magnitude smaller than the one observed by STAR.
We find the important result to have a quite good prediction for the $v_1(y)$ of $D$ mesons 
once the in medium HQ quark interaction able to correctly predict both $R_{AA}$ and $v_{2,3}(p_T)$ at RHIC and LHC is considered and the bulk longitudinal distribution is tuned to correctly predict the $v_1(y)$ for charged particles.
The absence of only one of these two features, tilted bulk and non-perturbative HQ interaction, would lead to a $v_1(y)$ much smaller than the one observed and comparable with the one of the charged particles.
Therefore, the very large $v_1(y)$ is indeed a further confirmation of the non-perturbative dynamics and in the future more precise measurement may contribute to the determination of the space-diffusion coefficient, in particular by Bayesian analysis \cite{Xu:2017obm}. 
We have also pointed out that the formation time of $v_1(y)$ for the charm quarks is quite short being of about only 2 fm$/c$.
This should imply that it is more sensitive to the HQ transport coefficient at the initial high temperature phase providing complementary information with respect to $v_2$ that is known to be more sensitive to the 
interaction at lower temperature close to $T_c$, as discussed in \cite{Das:2015ana}.

The other aspect studied is the splitting of $v_1(y)$ for  $\Dz$ and $\Dzb$ due to the electromagnetic 
fields. We have found that this is consistent with the
prediction at RHIC energy, but due to the current experimental data error bars would also be consistent with zero.
On the other hand the splitting at LHC energy appears to be quite large experimentally even if with still large error bars, but at present the theoretical models based on electromagnetic field profile assuming a medium in equilibrium at constant electric conductivity
is not able to account for it. However, it is likely that at LHC energies such an assumption is significantly inappropriate. 
In this respect it has to be considered that
modellings like the one employed here and in most of the current literature predicts a quite similar
magnetic fields at both RHIC and LHC,  see Fig. \ref{fig:EMF_rhicb9}.
On the other hand at LHC energy the initial maximum value expected in the vacuum is about a factor 50 larger than the one in a medium at the assumed $\sigma_{el}$, while such a gap is instead only about a factor of $3-4$ at RHIC energy. It is likely that at least a study of the electromagnetic field under a medium with a quickly varying electric conductivity is necessary, as well as, a study of the possible effects of anomalous chiral conductivity.

\begin{acknowledgements}
The authors appreciate very useful discussions with Gabriele Coci, Marco Ruggieri, Andrea Dubla and Ilya Selyuzhenkov.
L.O. is financially funded by the Alexander von Humboldt-Stiftung and acknowledges support from the Deutsche Forschungsgemeinschaft (DFG) through the grant CRC-TR 211 `Strong-interaction matter under extreme conditions'. V.G. and S.P. acknowledge the support of linea di intervento 2 DFA-Unict for the HQCDyn project.
\end{acknowledgements}



\bibliography{References}

\end{document}